%% file: CHI2026_CameraReady.tex
\documentclass[sigconf]{acmart}
\AtBeginDocument{%
  }

\copyrightyear{2026}
\acmYear{2026}
\setcopyright{cc}
\setcctype{by}
\acmConference[CHI '26]{Proceedings of the 2026 CHI Conference on Human Factors in Computing Systems}{April 13--17, 2026}{Barcelona, Spain}
\acmBooktitle{Proceedings of the 2026 CHI Conference on Human Factors in Computing Systems (CHI '26), April 13--17, 2026, Barcelona, Spain}
\acmPrice{}
\acmDOI{10.1145/3772318.3791679}
\acmISBN{979-8-4007-2278-3/2026/04}

\usepackage{booktabs}
\usepackage{tabularx}
\usepackage{subcaption,acmart-taps}
\usepackage{graphicx}  
\usepackage{array}
\usepackage{longtable}
\usepackage{lscape}
\usepackage{enumitem}
\usepackage{geometry}
\usepackage{multirow}
\usepackage{colortbl}
\usepackage{xcolor}
\usepackage{tikz}
\usepackage[normalem]{ulem}
\usetikzlibrary{calc}
\newcommand{\newtext}[1]{\textcolor{black}{#1}}

\begin{document}

\title [Feeling the Facts] {Feeling the Facts: Real-time Wearable Fact-checkers Can Use Nudges to Reduce User Belief in False Information}

\author{Chitralekha Gupta}
\authornote{Both authors contributed equally to this research.}
\email{chitralekha@nus.edu.sg}
\orcid{0000-0003-1350-9095}
\affiliation{%
  \institution{National University of Singapore}
  \country{Singapore}
}

\author{Nadia Victoria Aritonang}
\authornotemark[1]
\email{e1505949@u.nus.edu}
\orcid{0009-0005-6394-9445}
\affiliation{%
  \institution{National University of Singapore}
  \country{Singapore}}

\author{Dixon Prem Daniel Ranjendran}
\email{dixon@nus.edu.sg}
\orcid{0000-0001-9794-4435}
\affiliation{%
  \institution{National University of Singapore}
  \country{Singapore}
}

\author{Valdemar Danry}
\email{vdanry@media.mit.edu}
\orcid{0000-0001-5225-0077}
\affiliation{%
 \institution{MIT Media Lab}
 \city{Boston}
 \country{USA}}

\author{Pattie Maes}
\email{pattie@media.mit.edu}
\orcid{0000-0002-7722-6038}
\affiliation{%
  \institution{MIT Media Lab}
  \city{Boston}
  \country{USA}}

\author{Suranga Nanayakkara}
\email{suranga@ahlab.org}
\orcid{0000-0001-7441-5493}
\affiliation{%
  \institution{National Univerisity of Singapore}
  \country{Singapore}}

  \begin{teaserfigure}
  \centering
    \includegraphics[width=0.99\textwidth]{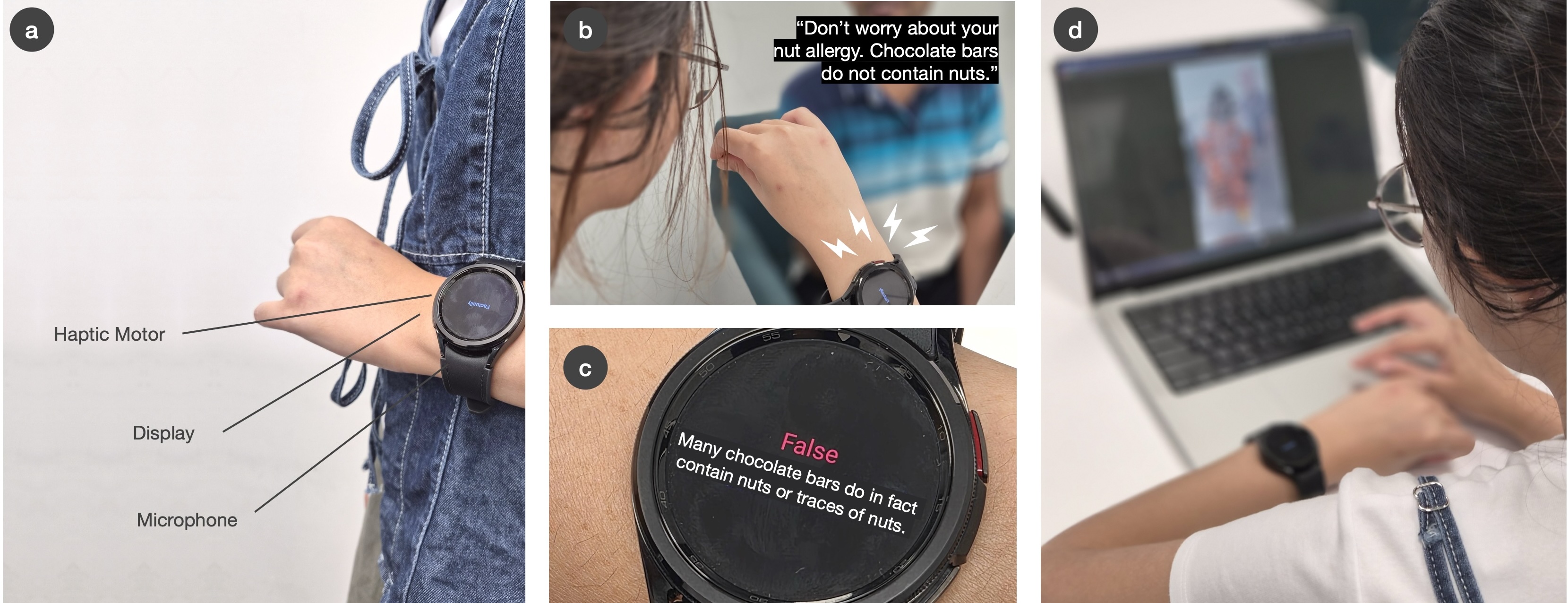}
     \caption{(a) The main components of the wearable fact-checker, ``Fact-Nudger'', (b) Fact-Nudger gives a subtle haptic vibration when a false claim is detected in an everyday setting, (c) A brief glanceable overview text appears on the watch when a false claim is detected, (d) A scene from the user study of this paper - the participant watching videos on a laptop, and the watch gives a haptic nudge when a false claim occurs in the video.}
    \Description{This figure consists of four parts: (a)  A person wearing the fact-checker called Fact-Nudger integrated with a smartwatch on their wrist. Three arrows pointing to different parts of the watch indicating Haptic motor, Display, and Microphone, (b) Fact-Nudger being used in an everyday setting where two people are having a conversation, one of them saying a false claim "Dont worry about the nut allergy. Chocolate bars do not contain nuts", and the other person wearing the fact-nudger gets a vibration or a haptic nudge on their smartwatch, (c) a zoomed in view of the watch screen that displays the text "False: Many chocolate bars do in fact contain nuts or traces of nuts", (d) Fact-Nudger being used by a person while they watch a video on a laptop.}
    \label{fig:teaser}
  \end{teaserfigure}

\renewcommand{\shortauthors}{Gupta et al.}

\begin{abstract}

Misinformation \newtext{can spread} rapidly in everyday conversation, where pausing to verify is not always possible. We envision a wearable system that bridges the timing gap between hearing a claim and forming a judgment. It uses ambient listening to detect verifiable claims, performs rapid web verification, and provides a subtle haptic nudge with a glanceable overview. A controlled study (N=34) simulated this approach and tested against a no-support baseline. Results show that instant, body-integrated feedback significantly improved real-time truth discernment and increased verification activity compared to unsupported fact-checking. However, it also introduced over-reliance when the system made errors, i.e.~failed to flag false claims or flagged true claims as false. We contribute empirical evidence of improved discernment alongside  insights into trust, effort, and user–system tensions in verification wearables.
\end{abstract}




\maketitle

\section{Introduction}
\newtext{Misinformation often occurs in everyday real-time interactions, such as} in meetings, hallway chats, classrooms, and calls—fast, real-life settings where pausing to verify can be costly. Decades of work show how misinformation can shape judgments even outside social media: Other people's reactions or preferences in daily life can subtly skew perceptions of what is true \cite{yang2023swaying}. For instance, cable-news framings affected COVID-19 behaviors and health outcomes \cite{bursztyn2020misinfo,simonov2020fox} where misperceptions correlated with which media outlet people were paying attention to \cite{jamieson2020media}. Interpersonal channels amplify these effects; beyond feeds and timelines, rumors and misbeliefs travel through everyday talk, meetings, and personal messaging networks \cite{friggeri2014rumor,vosoughi2018science}. In such moments, the friction to look things up means the conversation often moves on before verification catches up.
 
At the same time, cognitive science explains why these claims ``stick''. People lean on false or oversimplified intuitive heuristics when time/attention are scarce \cite{tversky1974heuristics}; repetition increases perceived truth \cite{pennycook2018prior}; and even after correction, discredited information can continue to guide people's thinking unless interventions are well-timed and well-designed \cite{johnson1994,lewandowsky2012, pennycook2021psychology, ecker2022psychological}. Recent work refines dual-process accounts by showing how ``feelings of right or wrongness'' can trigger shifts from intuitive to analytic processing and how individual differences in attuning to such  feelings can predict susceptibility to believing false information \cite{thompson2011intuition, pennycook2019lazy,bago2020fake}. While research has shown that ``accuracy prompts'', that is, paying attention to the accuracy of a statement, can help, they still require users to notice a claim, remember to care about accuracy, and take action \cite{pennycook2020,pennycook2021shifting, pennycook2022accuracy}. 

Two complementary lines of evidence suggest a different path. First, subtle, body-integrated cues (ambient displays, peripheral/haptic feedback) have been shown to shape attention and behavior with lower interruption costs than explicit pop-ups \cite{adams2015mindless,okeke2018good,signal2023}. Second, just-in-time adaptive interventions (JITAI) argue for micro-timed support at the very moment a decision is being formed \cite{intille2004ubiquitous,nahum2016just}. 
Yet ``always-listening'' assistance raises tensions between context-appropriate vs.\ intrusive feedback, helpful feedback vs.\ over-reliance, quality of feedback vs.\ latency to deliver that quality, and how user agency/privacy is preserved during ambient audio sensing \cite{lau2018alexa,malkin2019privacy}. 

Meanwhile, modern large language models (LLMs) can now combine speech recognition, claim-spotting, retrieval, and rapid synthesis—surfacing sources in seconds. Tool-augmented LLMs can browse the web and make citations \cite{nakano2021webgpt}, retrieve information on demand \cite{lewis2020retrieval}, interleave reasoning about the information with actions \cite{yao2023react}, and self-decide when to call tools \cite{schick2023toolformer,press2022measuring}. In misinformation specifically, claim-spotters find check-worthy statements in debates and speeches \cite{hassan2015detecting,hassan2017claimbuster, arslan2020benchmark, guo2022survey}, and new multimodal pipelines are emerging for video fact-checking and explanation \cite{niu2025pioneering}. Recent evidence even suggests dialogic AI can durably reduce conspiratorial beliefs, hinting at scalable, personalized debunking \cite{costello2024durably}. 

We explore whether a body-integrated \emph{wearable}, just-in-time scaffold can bridge the timing gap between hearing a claim and forming a judgment. We envision a system that listens in the background, detects potentially verifiable claims, performs a quick web verification, and, if signals warrant, issues a subtle haptic nudge, and a glance to the watch reveals a more detailed overview about the statement in question. By favoring ``verification-promoting cues'' over hard verdicts, this approach aims to trigger metacognitive checks without overstepping user agency, potentially building more reflection-oriented alternative pathways over continued use. Our approach builds on prior wearables that support reasoning about potentially misleading information \cite{danry2020wearable}. Within this framing, assistive augmentation \cite{tan2025assistive} provides a lens to view wearables not only as corrective tools but also as companions that strengthen reflective cognitive processes across diverse everyday contexts.

\newtext{Recent systems show the growing interest in assistive fact- verification tools. Wu et al.'s Factually \cite{wu2025factually} demonstrate how the concept of fact-check cues through a wearable could be useful during conversational misinformation, while Lee et al.~\cite{lee2024digital} develop a web-based pop-up for generative AI chatbots to subtly remind users to fact-check AI-generated outputs. Emerging open-source prototypes, such as the Gemini-Fact-Checker\footnote{https://github.com/VictoriousWealth/gemini-fact-checker.git}, illustrate ongoing efforts to build lightweight pipelines for automated claim verification. In this work, we closely examine how such body-integrated, just-in-time fact-check nudges delivered through wearables can affect human judgment and verification behavior.}
 
We report a controlled study (\(N=34\)) in which participants viewed pre-fact-checked videos with and without simulated wearable nudges. We test two hypotheses: 
\textbf{H1}: A real-time wearable fact-checker amplifies truth discernment compared to no wearable support. 
\textbf{H2}: Body-integrated, instant feedback yields higher verification activity (and perceived benefit) than delayed, self-initiated fact-checking (e.g., manual Google searches). We also explore over-reliance under system errors (eg.~false positives) and interaction of truth discernment with prior beliefs, topic involvement, and openness to changing one’s mind. 

Results show that nudges improved truth discernment and encouraged verification behaviors, but also introduced over-reliance risks when system errors occurred. These findings foreground design questions about calibration, such as communicating AI uncertainty, 
and ethical guardrails for ambient listening. In summary we contribute with: 
\begin{enumerate}
    \item A novel concept operationalizing subtle haptics with glanceable micro-explanations for in-situ truth discernment;
    \item Evidence that body-integrated, instant feedback improves discernment and verification behaviors relative to no support, alongside a failure-mode analysis (over-reliance under false positives and negatives);
    \item Insights on perceptions of trust, effort, and situational use, highlighting tensions between reassurance and skepticism, reflection and immediacy, and user agency and system control. 
\end{enumerate}
\newtext{}

\section{Related Work}
\subsection{Why Timing Matters for Correction}

When we encounter misinformation, our minds don't simply file it away as "unverified" or ``wrong''. Instead, false claims quickly weave themselves into our mental models of how the world works. Once embedded, these misconceptions prove remarkably stubborn. Even when we later learn that information was wrong—and even when we consciously remember the correction—the original falsehood continues to shape our reasoning and judgments—a phenomenon known as the "continued influence effect" \cite{johnson1994,lewandowsky2012,walter2020meta}. Each repetition compounds the problem. When we encounter familiar claims, our brains process them more fluently, mistaking this ease for truth even when we possess contradictory knowledge \cite{fazio2015knowledge,pennycook2018prior}.

To overcome this effect, simply correcting the false information is not enough, research has shown that people need coherent alternative explanations that fill the explanatory void left behind in order to remove the continued influence of the misinformation \cite{ecker2010explicit,johnson1994,rich2016continued,ithisuphalap2020does}.

Cognitive resources further constrain how well corrections can take hold. Under time pressure or distraction, we default to accepting what feels intuitively right rather than engaging in effortful verification \cite{pennycook2019lazy,bago2020fake}. For example, Sanderson et al. \cite{sanderson2023listening} examined corrections delivered while participants drove in a simulator and found that distraction significantly reduced correction effectiveness; a single correction was entirely ineffective under load, though multiple repetitions helped counter reliance on misinformation. Similarly in information environments, where the information is mixed with social or entertainment content, such as in a social media feed, or at a coffee break, attention towards accuracy is less present. These fragmented, attention-divided contexts lead to inattentive processing of source information, making us more vulnerable to misinformation's influence \cite{pearson2021sources}. When people navigate these mixed-context environments, they simply don't pause to consider accuracy \cite{pennycook2021shifting}.


These dynamics converge on a crucial insight: corrections succeed best when they arrive immediately, with sufficient explanations, right before repetition builds familiarity and while cognitive resources remain available for integration. Interventions that surface precisely when beliefs form stand the greatest chance of redirecting automatic acceptance toward deliberate reflection.


\subsection{Automated Fact-Checking and Interaction Techniques}


Researchers have explored multiple strategies to mitigate misinformation directly on information platforms. Lightweight prompts that encourage users to reflect on accuracy before posting have been shown to reduce sharing of false content in large-scale experiments \cite{pennycook2022accuracy,pennycook2021shifting}. Similarly, interventions that insert friction at the moment of sharing, such as pop-up reminders or interface nudges, have been shown to decrease willingness to spread misinformation \cite{jahanbakhsh2021exploring}. Platforms have also deployed debunking labels and banners that attach human-annotated fact-checking results to claims. While these labels can reduce belief in false content, repeated exposure can lead to ``label fatigue'' \cite{clayton2020}. More recent approaches draw on social non-specialist annotation, with community-driven notes surfacing contextual information alongside disputed claims \cite{allen2022birds}, while research prototypes have explored overlaying credibility signals or citations directly into video players to support lateral reading as content is consumed on platforms like Youtube or TikTok \cite{hughes2024viblio,hartwig2024adolescents}. Automated fact-checking pipelines extend these ideas further, aiming at detecting false claims in real time and retrieving candidate evidence \cite{hassan2017claimbuster,setty2024livefc}, though latency, accuracy and robustness remain open challenges. More recently, large-scale experiments with AI dialogues demonstrate the potential for durable belief change: Costello et al. \cite{costello2024durably} found that personalized GPT-4 conversations reduced conspiracy beliefs significantly. Their follow-up study, revealed that factual, reasoning-based counterarguments—rather than persuasion alone drive these effects \cite{costellojust}, highlighting the importance of evidence in AI-mediated debunking.

Despite these advances, existing interventions share critical limitations. They depend on users’ willingness to pause and attend to a screen, often target sharing rather than in-the-moment belief formation, and rarely arrive fast enough to interrupt the initial encoding of misinformation. Crucially, much misinformation also travels through off-screen channels such as word-of-mouth and everyday conversation, where labels, prompts, or overlays cannot reach \cite{allport1947psychology,coronel2020investigating}. This leaves a gap for interventions that are not screen-bound but instead integrate into daily life, capable of providing subtle, just-in-time cues during claim uptake. Our work explores one such direction through wearable fact-checking nudges.

\subsection{Wearables, Peripheral Cues, and Just-in-Time Nudges}

Just-in-time adaptive interventions (JITAIs) aim to deliver micro-timed support at the moment it is most likely to influence a decision. Originally developed in health contexts, JITAIs focus on matching the right type of assistance to the right moment, minimizing disruption while maximizing receptivity \cite{nahum2016just,klasnja2015microrandomized}. Recent work pushes this further: Haag et al. \cite{haag2025last} demonstrate that large language models can both detect opportune intervention windows and generate personalized nudges, suggesting that AI can expand the adaptivity and scalability of JITAI frameworks. For misinformation, this orientation toward timeliness is critical, since beliefs often solidify before traditional corrections arrive, and since explanations can be tailored to the users unique thinking style and prior knowledge.

HCI research on notifications reinforces that the \textit{when} and \textit{how} of intervention delivery determine its effectiveness. Studies show that poorly timed or obtrusive prompts impose high interruption costs, while lightweight, peripheral cues are more likely to be noticed and acted upon without derailing primary tasks \cite{pielot2014situ}. This has led researchers to explore ambient modalities such as vibrotactile feedback and subtle visual signals, which can convey information in parallel to ongoing activity. Tactons, structured haptic patterns, and indirect illumination have all been shown to communicate cues with low cognitive demand and high immediacy \cite{pousman2006taxonomy,brewster2004tactons,hoggan2009audio,pohl2016scatterwatch}. 

Social acceptability is equally important for interventions meant to operate in everyday life. Studies consistently find that wrist-worn devices, such as smartwatches, are perceived as more discreet and acceptable than conspicuous wearables like head-mounted displays \cite{koelle2020social}. Ethnographic accounts of smartwatch use further show how people naturally weave glances and vibrations into conversation without breaking social flow \cite{pizza2016smartwatch}. Together, this body of work positions the wrist as a socially compatible site for subtle nudges: a place where timely, low-friction cues can prompt reflection without demanding a pause or shift of setting. Building on these insights, our study evaluates a wearable fact-checking prototype that delivers haptic nudges paired with glanceable explanations to support truth discernment in the moment.


\subsection{AI-Augmented Reasoning and Towards Wearable Just-in-time Critical Thinking Support}
Studies of fallacies and reasoning highlight how people struggle to detect flawed arguments and how targeted interventions can strengthen critical thinking. Neuman and Weizman \cite{neuman2003role} showed that students’ ability to identify informal fallacies depends heavily on how argumentative texts are represented in the cognitive system, with deeper structural representations linked to better fallacy detection. Hruschka and Appel \cite{hruschka2023learning} tested a brief intervention and found that teaching participants about informal fallacies improved their ability to identify such arguments a week later, which in turn enhanced their discernment between real and fake news. Extending these insights into human–AI interaction, Danry et al. \cite{danry2023don} demonstrated that AI systems framed as questioners—engaging users in Socratic-style reasoning—improved logical discernment of divisive statements more effectively than simply providing causal explanations. These findings suggest that strengthening fallacy recognition, whether through cognitive training or AI-mediated questioning, can scaffold reflective reasoning and help counter the intuitive acceptance of misleading claims.

Literature reveals converging insights across cognitive science, intervention design, nudging theory, and wearable computing. Misinformation persists not only because of the nature of the claim but also due to human cognitive limitations, attentional constraints, and contextual frictions in everyday interactions. Interventions that are timely, subtle, and supportive of reflective reasoning—whether delivered as haptic nudges, podcast cues, AI dialogues, or Socratic questioning—offer promising avenues for promoting truth discernment without undermining user autonomy. This integrated body of work motivates the exploration of wearable, just-in-time scaffolds that can bridge the critical timing gap between encountering a claim and forming a judgment.



\section{Wearable Fact-checker ``Fact-Nudger'' Prototype Design}

\begin{figure*}
    \centering
    \includegraphics[width=0.8\linewidth]{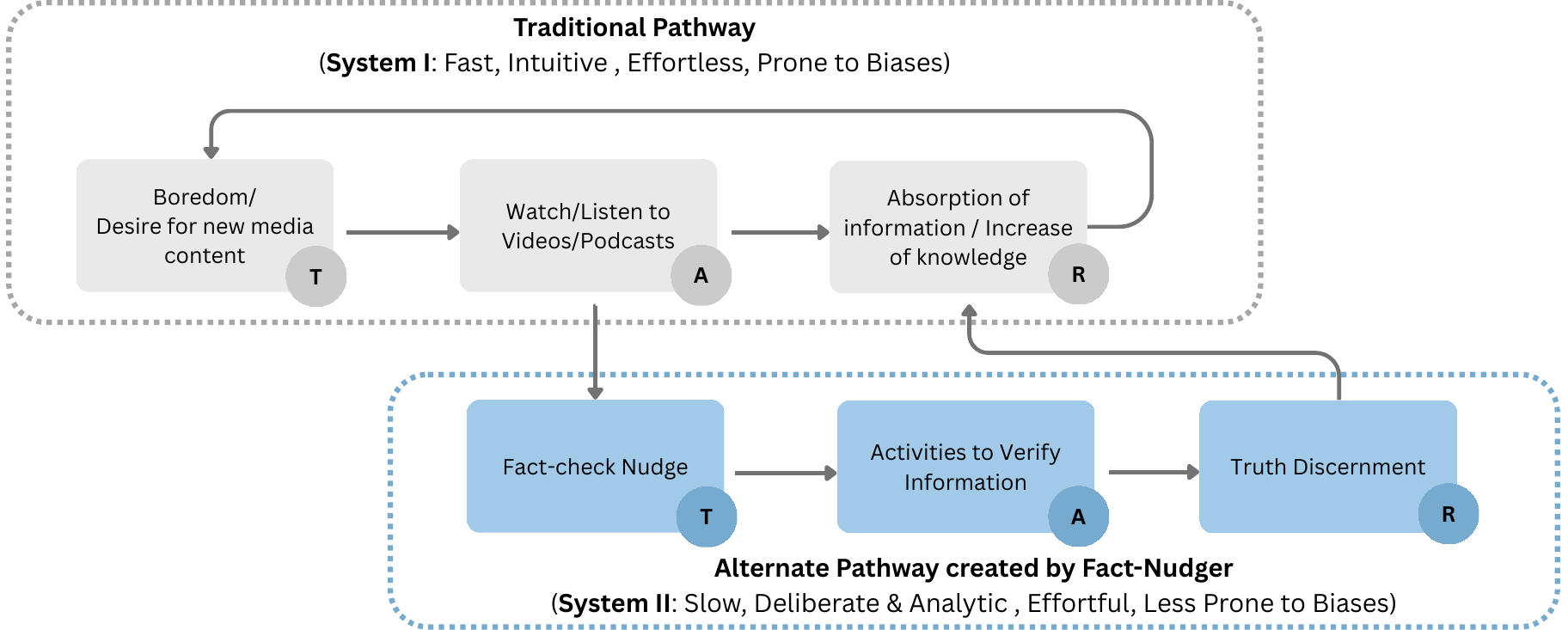}
    \caption{Overview: According to the dual-process theory, traditional pathway of information uptake (top) consists of triggers (T) such as curiosity leading to effortless engagement with new content (action (A)) and perceived knowledge gain (reward (R)), reinforcing fast, intuitive System 1 processing. Fact-Nudger (bottom) introduces subtle haptic nudges as alternative triggers, disrupting this loop and encouraging reflection and verification activities that engage slower, deliberate System 2 reasoning.}
    \Description{A flowchart with two pathways, Traditional and Alternate. The Traditional Pathway loops the following sequence: 1) Boredom/Desire for new media content, 2) Watch/Listen to Videos/Podcasts, 3) Absorption of information / Increase of knowledge. The Alternate Pathway starts from the Watch/Listin to Videos/Podcasts step in the Traditional Pathway and is followed by the sequence of: 1) Fact-check Nudge, 2) Activities to Verify Information, 3)Truth Discernment. After that, the Alternative Pathway converges back to the Absorption of information / Increase of knowledge step in the Traditional Pathway.}
    \label{fig:framing}
\end{figure*}

To investigate how a wearable real-time intervention would affect human behaviour of fact-verification and information processing, we developed Fact-Nudger, a wizard-of-oz prototype that simulates the experience of wearing a fact-checker while consuming media content. Fact-nudger sends a subtle haptic nudge along with a brief explanatory text whenever a false claim occurs in a pre-curated audio-visual media. Through this prototype, we study if these just-in-time subtle nudges can disrupt the intuitive information processing pathway (System 1) and nudge one towards the more analytic and reflective reasoning pathway (System 2) resulting in an improvement in truth discernment (Figure \ref{fig:framing}).


\subsection{Smartwatch as a Wearable Fact-checker}

We used a smartwatch that emits haptic feedback to nudge users when potential false information is presented. We chose the smartwatch form-factor due to its social acceptability as a day-to-day wearable and the discreetness of its haptic vibrations \cite{pizza2016smartwatch,whitmore2024improving}. A gentle buzz from a watch is unlikely to demand the user's full attention, while also allowing multi-tasking. Moreover, the widespread usage of smartwatches promotes the potential of deploying the system as a downloadable app, making the system accessible for people who already own a smartwatch.

The haptic feedback is synchronized with the misinformation, i.e.~as soon as a false claim occurs in the video, the haptic nudge is sent to the user\newtext{, i.e.~the vibration on the watch was triggered immediately after a false claim occurs}. In addition to haptic feedback, the watch also displays a brief textual explanation about the false claim. The text is timed to display at the same time as the watch vibration and disappears after a few seconds. This text is intended to give a quick explanation of why the statement they just heard was false. 

The android WearOS app was designed in AndroidStudio and deployed on a Galaxy Watch6 Classic (BLCR) SM-R950. 

\subsection{Video Media}
We presented information and misinformation through video clips curated from an existing fact-checked dataset~\cite{niu2025pioneering}. The clips were stitched into a continuous feed with bottom-to-top scrolling, simulating the experience of browsing social media. This design reflects how people typically encounter information—rapidly, across diverse domains, and without prior filtering. Using a pre-verified set of clips allowed us to run a controlled study on the wearable fact-checker’s effects on human behavior, while minimizing variability from real-world fact-checking.



We constructed two stimulus videos by sampling 16 short video clips from the dataset \cite{niu2025pioneering}. Each video contained eight clips: four presenting factually true claims and four presenting false claims, as verified in the original dataset annotations. Each clip averaged 35.75 seconds in duration, yielding a total runtime of approximately 4 minutes and 46 seconds per video. The clips spanned diverse everyday topics, including health, technology, animals, public figures, history, and geography. We deliberately avoided political or polarizing content to minimize confounding effects of prior biases. Examples of included claims ranged from ``\textit{Wearing socks to bed increases the blood circulation to your feet.}'' (true) to ``\textit{A man built an electric car that does not need to be charged as it converts radio frequencies to energy}'' (false). This range ensured that participants encountered a balanced mixture of plausible but verifiably correct and incorrect information. To keep the experience accessible and non-distressing, we prioritized general-interest clips and excluded content that might require extensive prior knowledge or elicit strong emotional responses. The complete list of claims in the video clips used in this study\footnote{Video 1: \url{https://shorturl.at/0x96W}\\Video 2: \url{https://shorturl.at/vae1p}} is provided in Table \ref{tab:claims} in Appendix.

\subsection{Prototype Interaction Flow}

\begin{figure*}[h]
    \centering
        \includegraphics[width=0.80\textwidth]{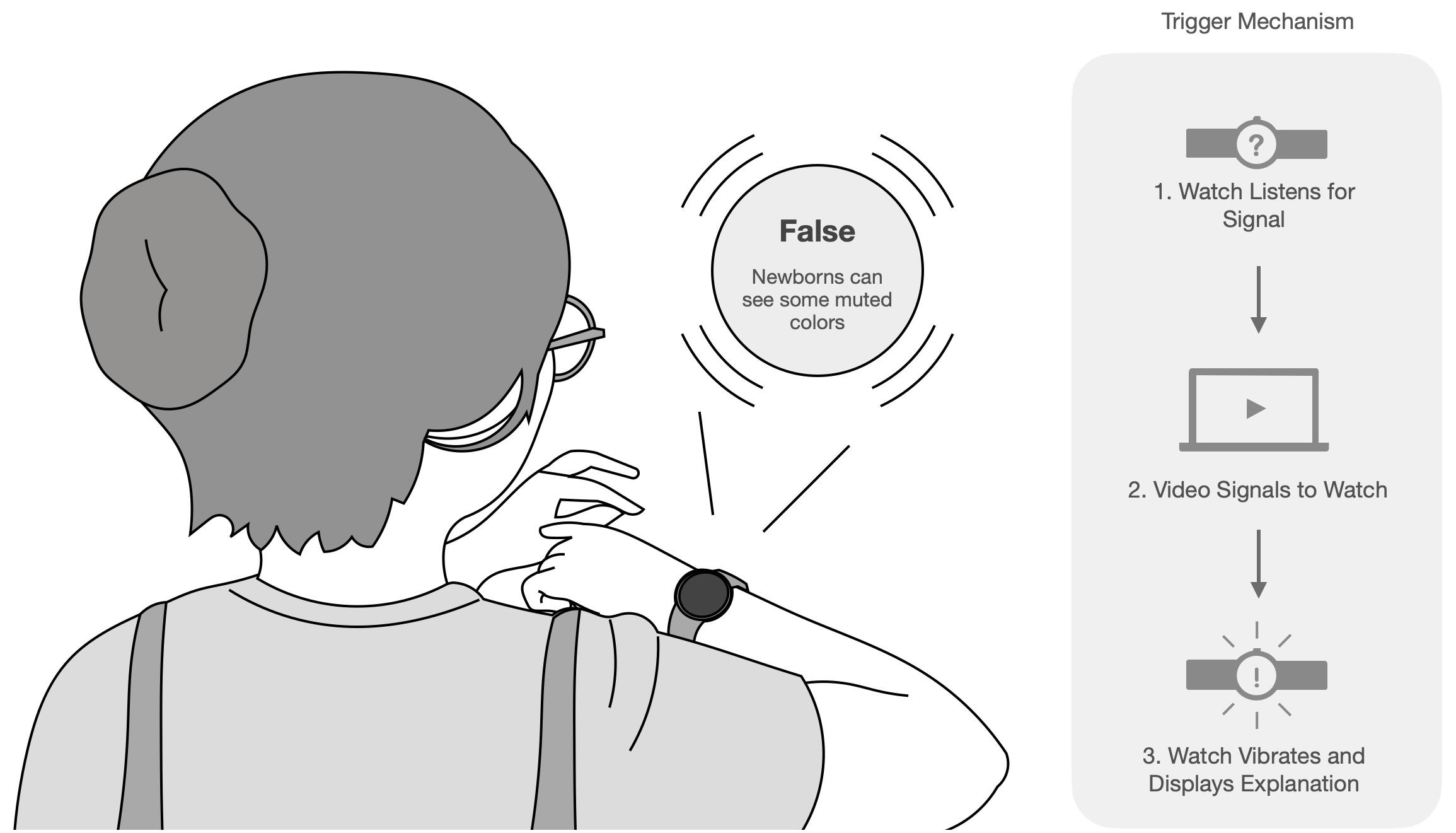}
    \caption{Prototype Interaction Flow\newtext{, that starts with the smartwatch listening for signals, the video sends signals to the watch when certain claims occur, and finally the watch vibrates and displays an explanation.}}
    \Description{The image depicts how the Fact-Nudger prototype works. An individual is wearing a smartwatch that is listening for signals. A laptop with the video player is sending a signals to the watch during at a marked timestamp. The watch vibrating, with text on its screen showing the text "False" and a claim it is explaining."}
    \label{fig:prototype_flow} 
\end{figure*}

We built a custom video player that sent timestamped cues to the wearable whenever a claim appeared. At each marked misinformation timestamp, the watch vibrated and displayed a short text explanation, simulating an AI-driven fact-checker (Figure~\ref{fig:prototype_flow}). This interaction loop continued as participants watched the video until playback ended.  

To emulate the imperfect accuracy of real AI systems, we benchmarked current video fact-checkers such as 3MFact (79.6\%)~\cite{niu2025pioneering}, FakeSV (79.31\%)~\cite{qi2023fakesv}, and FakingRecipe (79.15\%)~\cite{bu2024fakingrecipe}. Given that our study includes a small number of claims per video, we introduced \newtext{1-2} deliberate errors per video, either by omitting a nudge (false negative) or inserting an extra nudge (false positive). In both cases, system accuracy was at 87.5\%\footnote{In case of a false positive, i.e.~a true claim detected as false, (TP=4, FP=1, TN=3, FN=0) (TP=True positive, FP=False positive, TN=True negative, FN=False negative), resulting in accuracy ((TP+TN)/(TP+TN+FP+FN)) of the system to be 87.5\%. Similarly, in case of a false negative, i.e.~a false claim detected as true, TP=3, FP=0, TN=4, FN=1, resulting in the same accuracy.}, slightly higher than the state-of-the-art but still reflecting a realistically imperfect AI system. This design allowed us to study how participants responded to erroneous nudges.

\section{Experiments}
\subsection{Participants}
We recruited $N = 34$ participants (20 males, 14 females), who were university staff or students. Participants were predominantly between 25–34 years old (71\%), with fewer in the 18–24 (21\%), 35–44 (6\%), and 45-55 (3\%) ranges. All participants reported fluent proficiency in English and provided informed consent. Participants received 10 dollars for their participation. The study was approved by the National University of Singapore IRB committee.

\subsection{Study Design}
We employed a within-subjects experimental design with two conditions:
\begin{itemize}[leftmargin=10pt]
    \item \textbf{Treatment condition (wearable W)}: participants watched a compilation of short video clips (total video duration 4 min 46 seconds) containing pre-verified true and false statements while wearing fact-nudger. The watch provided a haptic nudge when potential misinformation was detected (accuracy 87.5\%).
    \item \textbf{Control condition (no-wearable NW)}: participants watched a comparable compilation of video clips of the same duration without wearable support.
\end{itemize}

\begin{figure*}
    \centering
    \includegraphics[width=\linewidth]
{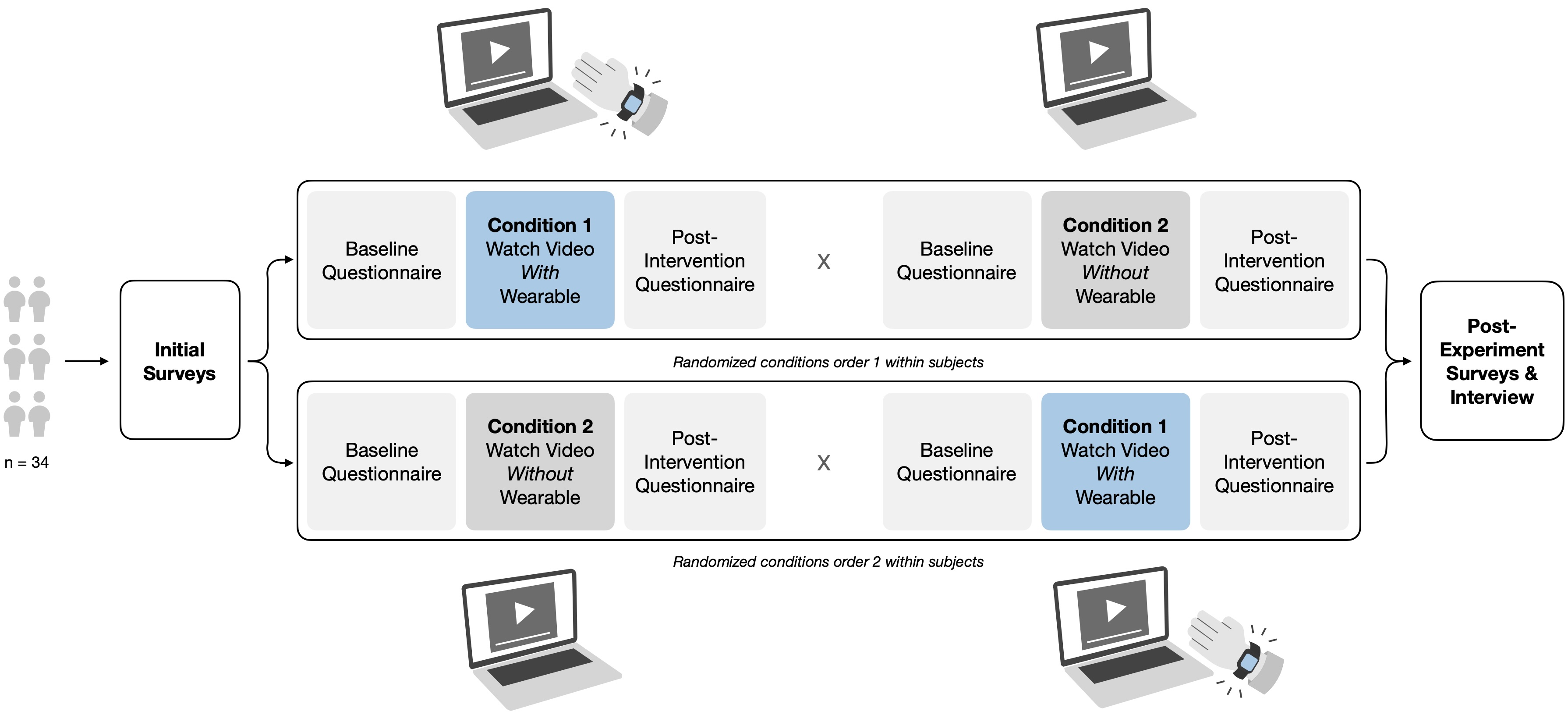}
    \vspace{-0.5cm}
    \caption{Overview of the experimental procedure.}
    \label{fig:studydesign}
    \Description{The flowchart depicts the study design procedure for two groups, each containing 17 participants, totaling 34 participants. The process begins with an initial questionnaires session. Participants are then split into two groups of 17 each. The first group's sequence includes baseline questionnaire 1, watching video 1 without wearable, post-intervention questionnaire 1, baseline questionnaire 2, watch video 2 with wearable, then post-intervention questionnaire 2. The bottom group follows a similar procedure but starts with video 1 with wearable and ends with video 2 without wearable. Both groups conclude with a qualitative Questionnaire and interview.}
\end{figure*}

The order of these conditions was counterbalanced across participants. Each video compilation included 8 short clips: \newtext{the first video compilation had 5 true claims, 3 false claims, and the second video compilation had 4 true and 4 false claims.} 
\newtext{Instruction to the participants before the control condition was, ``\textit{You will watch a video containing true and false statements. Please watch it as you would naturally. Feel free to pause, rewind, skip, or use web browser (e.g., Google, ChatGPT) at any time}''. And for the wearable condition, an additional instruction was given, ``\textit{You will wear a smartwatch while watching the video that has an AI system. The watch might vibrate and display a text message when a potential misinformation is detected.}''. The participants were not given any information about the accuracy of the system.} The study design is shown in Figure \ref{fig:studydesign}.

\newtext{The sessions were video recorded to capture participants’ interactions with the video (pauses, rewinds, and fast-forwards), their browser activity, and the count of glances at the smartwatch. During each session, a researcher manually logged these events, which were verified later again by reviewing the video recordings. The total duration of the study per participant on an average was 1 hour 8 minutes.}
\subsection{Questionnaires}
Participants completed a series of questionnaires at different stages of the study to capture their demographics, baseline verification tendencies, belief in specific claims before and after the intervention, and their experience with the intervention.
\subsubsection*{Initial Questionnaire: }
The initial questionnaire collected demographics (age, gender) and assessed participants’ verification behaviors. Participants rated how often they verify information across different media on a 7-point scale, reported their preferred verification methods, and identified barriers to verification. Additional items measured trust in AI-provided information, the most recent verification instance, and willingness to verify claims across domains (e.g., health, public figures, technology), which aligned with topics later shown in the videos. Finally, openness to updating beliefs was assessed using the Actively Open-Minded Thinking scale~\cite{stanovich2023actively}. A full overview of questionnaire items is provided in Appendix, Table ~\ref{tab:questionnaire}.

\subsubsection*{Baseline Questionnaire: }
To measure prior exposure and baseline belief in claims, participants evaluated ten statements (eight appearing in the following stimulus videos and two decoys). For each statement, they answered the following questions adapted from Pennycook and Rand \cite{pennycook2019lazy} for specific constructs: 
\begin{itemize}[leftmargin=10pt]
    \item \textbf{Prior exposure:} ``Have you seen or heard about this statement before?'' (yes / unsure / no)
    \item \textbf{Baseline Belief:} ``To what extent do you believe this statement?'' (7-point Likert scale - 1: Definitely False, 7: Definitely True)
    \item \textbf{Baseline confidence in belief:} ``How confident are you in your belief?'' (7-point Likert scale: 1: Not confident, 7: Very confident)
\end{itemize}
The list of claim statements for both videos are provided in Appendix Table \ref{tab:claims}.

\subsubsection*{Post-Intervention Questionnaire: }
After each video, participants answered the same belief and confidence questions for the corresponding claims adapted from Ecker et al. \cite{ecker2017reminders} measuring \textbf{post-intervention belief} and \textbf{post-intervention confidence in belief}, along with an additional question, ``How important is it for you to verify this claim?'' (rating for \textbf{care for the topic}). 
Participants also completed the 
NASA-TLX \cite{hart1988development} form to measure subjective cognitive workload of their experience in each condition.
\subsubsection*{Qualitative Feedback and Interview: }
\label{sec:qual}
Following both video-watching sessions, participants answered two open-ended questions in writing, reflecting on their experiences with and without the wearable system. Finally, a semi-structured interview was conducted to obtain more detailed feedback and suggestions.

\subsection{Measures and Hypotheses}
\label{sec:measures}

Our study evaluated the effectiveness of real-time wearable fact-checkers along two primary dimensions: 
(1) their ability to reduce belief in false claims (\textbf{H1}), and 
(2) their ability to promote verification behavior (\textbf{H2}). 
In addition, we explored how individual differences (e.g., topic care, Actively Open-Minded Thinking (AOT), and prior exposure to misinformation) and system errors moderate these effects. For measuring statistical significance of the differences, we first conducted Shapiro-Wilk to test if the distribution is normal. If it is, then we used parametric two-sided t-test, and if not we used non-parametric Wilcoxon signed-rank test.

\subsubsection*{\textbf{H1: Reducing Belief in Misinformation}}
\begin{itemize}[leftmargin=*]
    \item \textit{H1.1:} Real-time wearable fact-checker nudges will reduce user belief in false information. \\
    \textit{Measures:} Belief ratings across conditions (Wearable/No-Wearable, Baseline/Intervention, True Claims/False Claims), and changes in belief (Intervention -- Baseline)
    \item \textit{H1.2:} Real-time wearable fact-checker nudges will increase confidence in rejecting false information. \\
    \textit{Measures:} Confidence ratings across conditions (Wearable/No-Wearable, Baseline/Intervention, True Claims/False Claims). 
\end{itemize}

\subsubsection*{\textbf{H2: Promoting Verification Behavior}}
\begin{itemize}[leftmargin=*]
    \item \textit{H2.1:} Real-time wearable fact-checker increases user engagement with verification activities\\
    \textit{Measures:} Number of verification activities, i.e.~pauses, rewinds, web searches, watch glances. Glancing at the watch was considered as a verification activity in the wearable condition as the participants were informed before the intervention that the watch is going to buzz and display an explanation text when a potential misinformation occurred. Thus from the user's perspective, glancing at the watch was a verification activity. 
    \item \textit{H2.2:} Real-time wearable fact-checker increases the time spent on claim verification\\
    \textit{Measures:} Total time spent watching the videos (minus the duration of the videos, which is a constant across all participants).
\end{itemize}

\subsubsection*{\textbf{Additional Analysis}}
In addition to the core hypotheses, we examined several additional conditions:  
\begin{itemize}[leftmargin=*]
    \item \textit{Effect of AI errors:} Erroneous nudges (false positives/false negatives) may reduce the effectiveness of the intervention.\\
    \textit{Measures:} For wearable condition, change in belief across nudges for  (true positive), correct non-nudges (true negative), extra nudges (false positive), and missed nudges (false negative).
    \item \textit{Effect of care for the topic:} Wearable fact-checker nudges can reduce belief in false information more effectively if the user cares about the topic of the claim.\\ 
    \textit{Measures}: Relationship between care ratings and change in belief (Wearable/No-Wearable).
    
    \item \textit{Effect of AOT:} Individuals with higher AOT may respond more strongly to nudges, in terms of belief updating.\\
    \textit{Measures}: Relationship between AOT scores and change in belief (Wearable/No-Wearable).
    \item \textit{Effect of prior exposure:} Prior exposure to claims may not change the effectiveness of nudges in correcting belief.\\
    \textit{Measures}: Relationship between prior exposure (yes/no/unsure) to the claim before the study and change in belief (Wearable/No-Wearable)
    
\end{itemize}

\section{Results}
\subsection{Baseline verification habits and barriers}
Prior to the experimental intervention, we collected baseline data from all participants to gauge their pre-existing information verification habits, perceived barriers, and topic interests (All plots for this subsection are in Appendix).

\textbf{Primary Verification Methods.} When asked to select all methods they use to verify information, participants reported relying most heavily on Google/search engines (90\% of participants), establishing it as the dominant tool for independent fact-checking. This was followed by asking someone they trust (50\%) and using AI chat platforms (45\%). Notably, dedicated fact-checking websites (e.g., Snopes, Politifact) were used by a smaller portion of the cohort (35\%), and actively looking for platform misinformation labels was the least selected method (20\%), suggesting a lack of availability, visibility, or trust in these native platform interventions in this participant group.

\textbf{Frequency of Verification by Context.} The frequency with which participants verify information varies significantly depending on the platform or context. Verification behavior is most consistently reported for information encountered on online news articles and TV news, indicating a higher degree of inherent skepticism towards traditional media formats. Conversely, information absorbed through podcasts/audio content and, most notably, during casual conversations, is verified the least. This suggests that social contexts and ephemeral audio formats present significant hurdles for proactive verification.

\textbf{Barriers to Verification.} When ranking the most significant barriers to verifying information, \textit{Lack of time} emerged as the primary obstacle, highlighting the cognitive effort and interruption required by current verification methods. This was closely followed by \textit{Forgetting to check}, which highlights the need for proactive, rather than reactive, verification tools. Other notable barriers included simply \textit{Agreeing with the claim} (confirmation bias) and a lack of familiarity with the topic, while not knowing how to verify was a less significant factor for this group.

\textbf{Topic Interest.} Amongst the topics presented, participants were most interested in verifying information related to Health and Technology, topics with direct personal impact or rapid development. Interest was lower for topics like History, Public Figures, and Geography.

Thus, the baseline data reveals a participant group that is equipped with the knowledge of how to verify (primarily via search engines) but is hindered by practical constraints such as time and forgetfulness. This establishes a clear need for low-friction, real-time verification support systems, which our wearable fact-check nudges aim to provide.
\subsection{Wearable fact-checker nudges reduce user belief in false information (H1)}
\label{sec:H1quant}
\begin{figure*}[t]
    \centering
      \includegraphics[width=\textwidth]{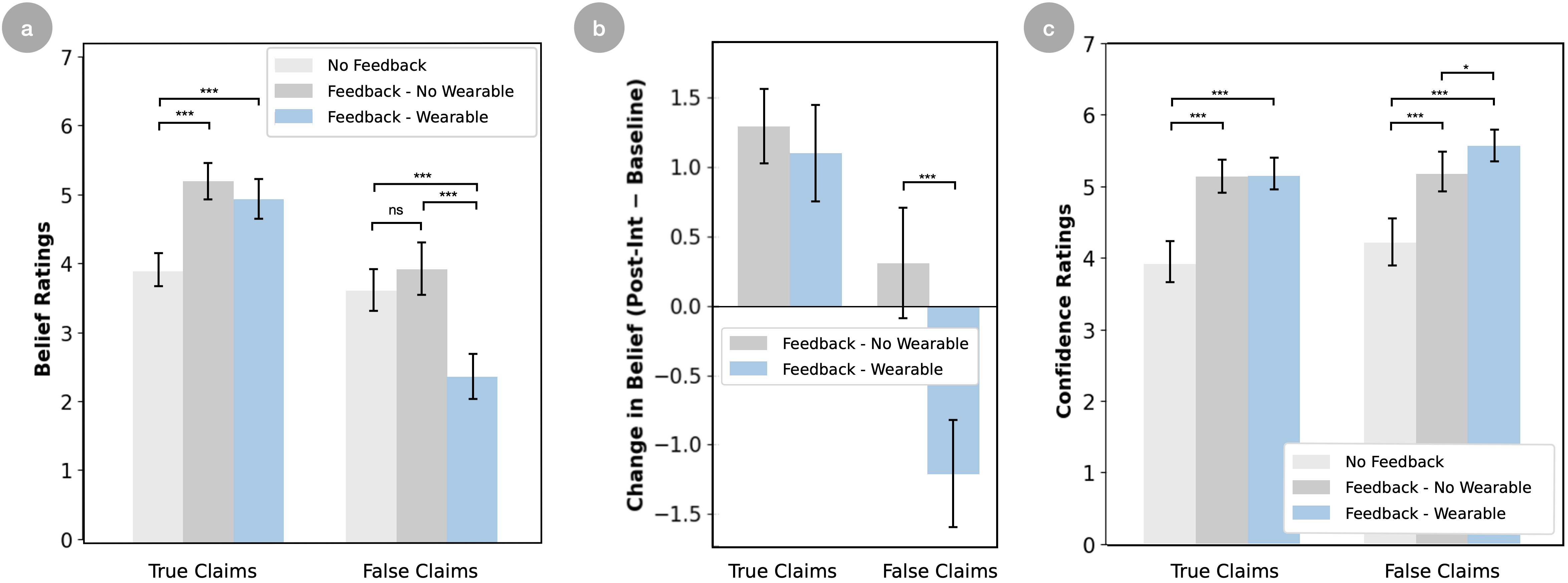}
    \caption{Results for H1: Reducing Belief in Misinformation: (a) belief ratings across all conditions, (b) change in belief (post study -- baseline; -ve belief change means reduction in belief after intervention), and (c) confidence ratings across conditions. Error bars represent 95\% confidence interval. Wilcoxon signed-rank test *p<0.05, **p<0.001, ***p<0.0001.}
    \Description{Figure (a): Belief Ratings
This bar chart compares average belief ratings for both true and false claims across the two groups and two time points: "Baseline" (before the intervention) and "Post-Intervention" (after the intervention).
For True Claims:
The baseline belief rating was around 4.0 for the no-wearable group and slightly lower, around 3.8, for the wearable group.
After the intervention, the belief rating for true claims increased significantly to around 5 for both groups.
This shows that both the wearable and no-wearable groups increased their belief in true claims, with a significant increase from baseline to post-intervention in both cases.
For False Claims:
The baseline belief rating for false claims was similar for both groups, around 3.5.
After the intervention, the no-wearable group's belief rating remained stable, staying around 3.5.
In contrast, the wearable group's belief rating for false claims decreased significantly, from around 3.5 at baseline to approximately 2.5 after the intervention.
Figure (b): Change in Belief
This chart shows the change in belief (post-intervention minus baseline) for both true and false claims. A positive change means belief increased, and a negative change means it decreased.
For True Claims:
Both the no-wearable and wearable groups showed a positive change in belief, with the wearable group's change being slightly larger, compared to the no-wearable group's change. There was no significant difference between the two groups' change in belief for true claims.
For False Claims:
The no-wearable group showed a very small positive change in belief, staying near zero.
The wearable group, however, showed a statistically significant negative change, with a drop in belief of approximately -1.25. This indicates a significant reduction in belief in false claims for the wearable group, which was not seen in the no-wearable group.
Figure (c): Confidence Ratings
This bar chart compares average confidence ratings for true and false claims.
For True Claims:
At baseline, confidence ratings were similar for both groups, around 4.
After the intervention, confidence ratings for true claims increased significantly in both the no-wearable and wearable groups, rising to around 5.5. This indicates both groups became more confident in their belief in true claims.
For False Claims:
At baseline, confidence ratings were similar for both groups, around 4.5.
After the intervention, the confidence in rating the false claims also increased significantly for both the groups. However, the confidence of the wearable group was significantly higher than the no wearable group post-intervention.}
    \label{fig:h1_results}
\end{figure*}

Figure \ref{fig:h1_results} shows that haptic nudges from real-time wearable fact-checker effectively reduces belief in misinformation while increasing confidence in rejecting false information. 
As shown in Figure \ref{fig:h1_results}(a), users wearing the device exhibited significantly lower belief in false claims post-intervention \newtext{($M=2.38, SD=1.82$)} compared to their baseline \newtext{($M=3.61, SD=1.76$)} 
\newtext{($W = 643.5, p < 0.0001, r = 0.56$, where, total number of false claims = 7; Number of pairs of comparison for false claims, n = 7 $\times$ 17 = 119, and degrees of freedom $df$ is n-1=118)}. Furthermore, belief scores in false claims post-intervention in the wearable condition \newtext{($M=2.38, SD=1.82$)} were also significantly lower than those in the no-wearable condition \newtext{($M=3.92, SD=2.09$)} 
\newtext{($W = 689.0, p < 0.0001, r = 0.66$, $df = 118$)} (Figure \ref{fig:h1_results}(a)). A Wilcoxon signed-rank test was used for both comparisons as the difference scores were not normally distributed (Shapiro-Wilk $p < .05$).
However, \newtext{between the two conditions, the participants} 
\newtext{did not show a significant difference} in belief in true claims post-intervention 
\newtext{($W = 3077.0, p = 0.24, r = 0.02$, and Shapiro-Wilk $p < .05$, where, total number of true claims = 9; number of pairs of comparison for true claims = 9 $\times$ 17 = 153, thus $df = 152$)}. Figure \ref{fig:h1_results}(b) further quantifies this result, showing a statistically significant reduction in belief for false claims among participants in the wearable condition \newtext{($M=-1.23,SD=2.16$)} compared to the no-wearable condition \newtext{($M=0.31,SD=2.18$)}. This difference was confirmed by a paired t-test (selected due to normal distribution of the differences Shapiro-Wilk $p = 0.088$), and showed a large effect 
\newtext{($t = 5.76, p < .0001$, Cohen's $d = 0.53$, $df = 118$)}. 
Complementing these findings, Figure \ref{fig:h1_results}(c) shows that users showed a higher confidence in rejecting false claims in the wearable condition \newtext{($M=5.72,SD=1.36$)} compared to no-wearable condition \newtext{($M=5.31,SD=1.47$)}. A Wilcoxon signed-rank test (used due to non-normality of the difference scores, Shapiro-Wilk $p = 0.018$) confirmed this result was statistically significant 
\newtext{($W = 1600.5$, $p = 0.023$, $df = 118$)}. 
This indicates not just reduced belief but increased confidence in identifying misinformation. Together, these results suggest that real-time nudging through wearable fact-checkers effectively counters misinformation while preserving belief in accurate information.





\subsection{Wearable fact-checker nudges increase verification activities in users (H2)}

\begin{figure*}[h]

    \centering
    \includegraphics[width=0.8\textwidth]{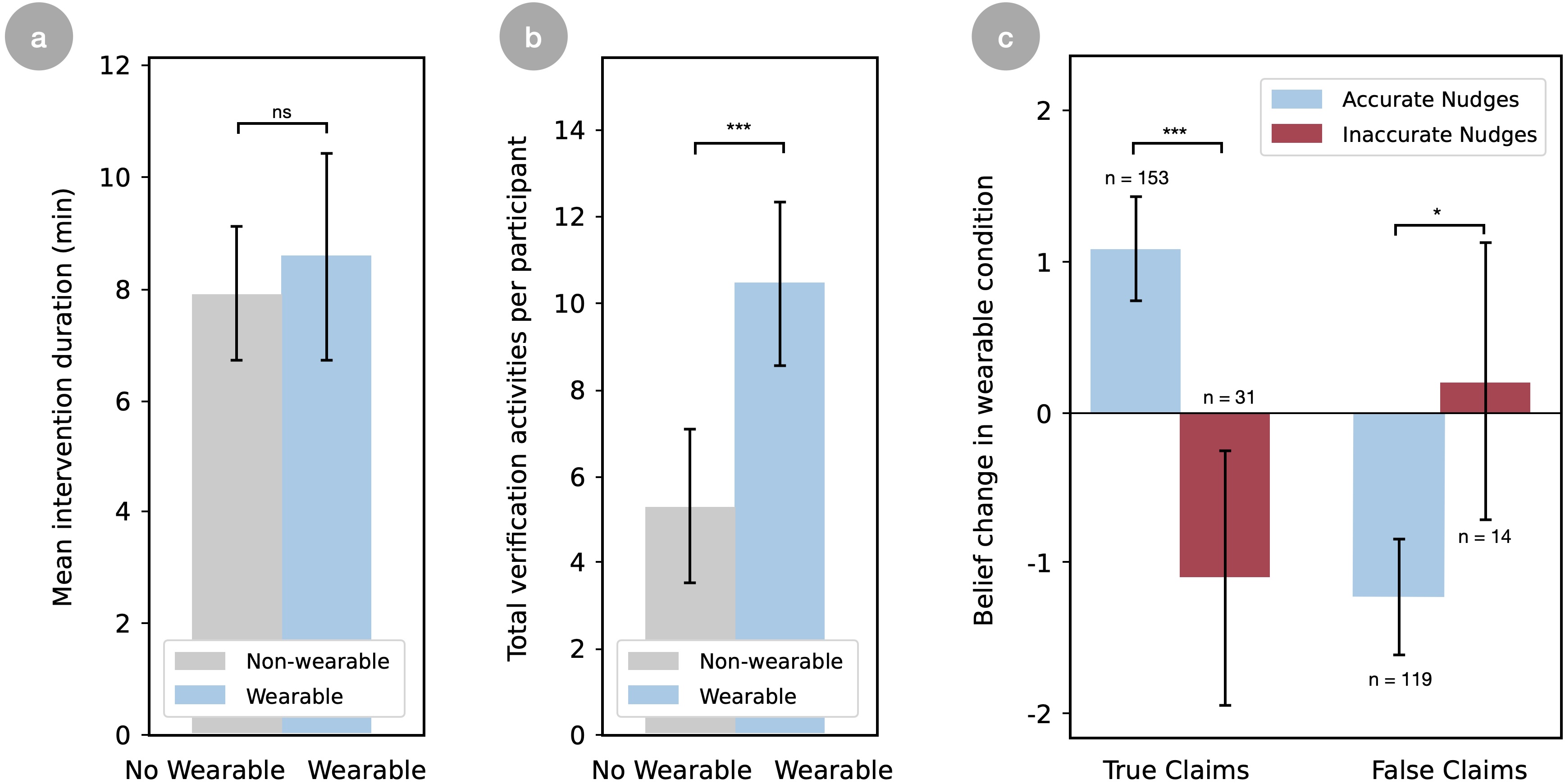}%
    \Description{Figure (a) The bar chart has two bars, each representing mean of total verification activities per participant (y-axis), ranging from 0 to 12 activities. The x-axis features two categories: No wearable and wearable. The no wearable bar is just about 5 activities. The wearable bar is slightly above 10 activities. The difference between the two groups is significant. Figure (b) The bar chart has two bars, each representing mean intervention duration in minutes. The y-axis is labeled mean intervention duration (minutes) and ranges from 0 to 10. The x-axis features two categories: No wearable and wearable. The no wearable bar is just about 8 minutes. The wearable bar is slightly above 8 minutes. (c) The bar chart displays the belief change in a wearable condition, comparing the effects of nudges on both true and false claims. The horizontal axis has four labeled categories: True claims without nudges (TN), true claims with nudges (FP), false claims with nudges (TP), and false claims without nudges (FN). The belief change for true claims without nudges is around 1.5 and is significantly taller than the belief change for true claims that incorrectly had nudges, which is around -1. The belief change for false claims with nudges are around -1.5, while the false claims that incorrectly did not have nudges is slightly above 0.}
    \caption{(a) Results for H2: Promoting Verification Behavior -- Mean intervention duration (minutes), (b) Results for H2: Promoting Verification Behavior -- Total verification activities per participant. Error bars represent 95\% confidence interval. Wilcoxon signed-rank test, ***p < 0.0001, ns is not significant. 
    (c) Impact of nudge accuracy on belief change in the wearable condition. The inaccurate nudges for true claims indicate false positives (FP) or extra nudges, and the inaccurate nudges for false claims indicate false negatives (FN) or the nudges that were removed. Error bars represent 95\% confidence interval. Mann Whitney U test *p < 0.05, ***p < 0.0001. }

\label{fig:h2_results}
\end{figure*}

As shown in Figure \ref{fig:h2_results}(b), participants using the wearable fact-checker engaged in a significantly higher number of verification activities (e.g., pausing, rewinding, web searching, glancing at the watch) \newtext{($M=10.47,SD=5.59$)} compared to those in the non-wearable condition \newtext{($M=5.26,SD=5.38$)}. A paired t-test confirmed this difference was statistically significant ($t = 5.769$, $p < 0.0001$, \newtext{where the number of participants for each condition, and hence number of comparisons, n = 34, and degrees of freedom $df$ is n-1=33}); the use of a parametric test was justified as the difference scores did not significantly deviate from normality (Shapiro-Wilk $W = 0.941$, $p = 0.06$). Glancing at the watch was an additional verification activity in the wearable condition, as mentioned in Section \ref{sec:measures}. However, 
Figure \ref{fig:h2_results}(a) shows that this additional activity did not significantly increase the average time (in minutes) invested in the verification process (Wilcoxon $W = 228.0$, $p = 0.93$, \newtext{$df = 33$}; Shapiro-Wilk normality test $p < 0.0001$ showing distribution not normal). This implies that the act of glancing at the watch provided a seamless, low-cost, immediate form of verification that supplemented, rather than replaced, other activities without imposing a significant time burden on the user.

\subsection{Impact of Erroneous Nudges on Belief Change}

The simulation of AI errors, i.e.~one extra nudge (false positive) or one missed nudge (false negative) per participant, showed a significant effect on user belief. As shown in Figure \ref{fig:h2_results} (c), false positive nudges (triggered for true claims) caused a substantial decrease in belief for accurate information. A non-parametric unpaired two-sided Mann-Whitney U test\footnote{The Mann-Whitney U test was used because the analysis compared two independent groups (n=153 for correct claims vs. n=31 for extra nudges, and n=119 for false claims vs n=14 for missed nudges) with unequal and small sample sizes, and the belief change scores violated the normality assumption.} confirmed this effect was statistically significant ($U = 3627.50$, $p < 0.0001$). 
Conversely, false negative errors (missing nudges for false claims) resulted in a failed correction, with belief change remaining near zero. This lack of correction was confirmed by a Mann-Whitney U test, which showed a statistically significant difference in belief change compared to the correctly nudged condition ($U = 507.50$, $p = 0.02$). 
This indicates that due to the false positives, people became significantly more skeptical towards true information implying that they relied heavily on the nudges to make a judgment. When a nudge was missed, the change in belief was small, meaning people stuck to their baseline instinct, and did not actively increase their belief in the claim because the watch did not buzz. These results show that there is a tendency of over-reliance on the buzzes from the watch, but there is also a tendency to fallback on natural instincts (and baseline fact-checking behaviour) when there is a missed nudge.

\subsection{Wearable Intervention Corrects Beliefs Consistently Across Care Levels}

As shown in Figure \ref{fig:topiccare}, there was no significant correlation between care score and belief change for false claims in the wearable condition ($r = -0.13, p = 0.26$). This indicates that the wearable fact-checking intervention may work consistently across different levels of topic relevance, effectively decoupling corrective outcomes from users' personal interest in the subject matter.

In contrast, a weak but statistically significant positive correlation was observed in the no-wearable condition ($r=0.19,p=0.025$). This counterintuitive finding suggests that without the wearable intervention, participants who cared more about a topic showed a slight increase in their belief in false claims. A possible explanation for this lies in the nature of passive video consumption. As noted by Sundar et al.~\cite{sundar2021seeing}, users may default to heuristic processing, where the visual authority of a video makes information seem more credible. It is plausible that, in the absence of a verification nudge, participants who cared more about the topics may have been more susceptible to this effect, relying on the video content without engaging in critical verification. Indeed, in several instances, this manifested as participants watched the video that had a professional look but had false claims, but did not seek external verification, allowing their pre-existing interest to amplify the message's persuasiveness.

\aptLtoX{\begin{figure}[h]
    \centering
        \includegraphics[width=\textwidth]{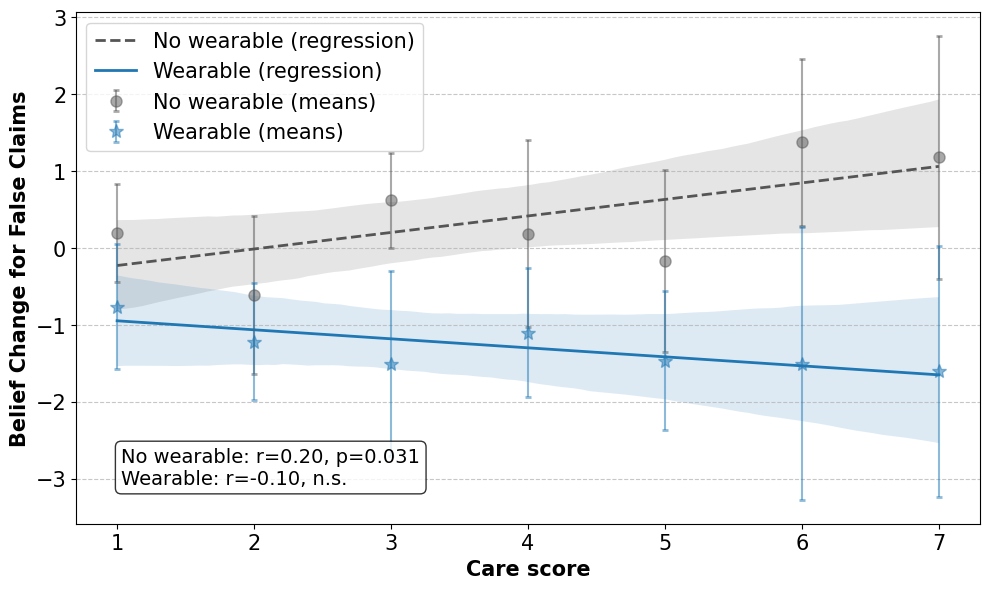}
        \vspace{-0.5cm}
        \caption{The effect of topic care on belief change (post study -- baseline; -ve belief change means reduction in belief after intervention) for false claims, by condition. Error bands represent 95\% confidence intervals around linear regression.}
        \label{fig:topiccare}
        \Description{The line graph displays the relationship between care scores and belief change for false claims, comparing groups with and without wearables. The x-axis represents the care score, ranging from 1 to 7, while the y-axis represents the belief change for false claims, from -3 to 3. There are two sets of data: one for no wearable and another for wearable. The no wearable shows a slight positive slope, while the wearable shows a slight negative slope.}
    \end{figure}
    \begin{figure}
        \centering
        \includegraphics[width=\textwidth]{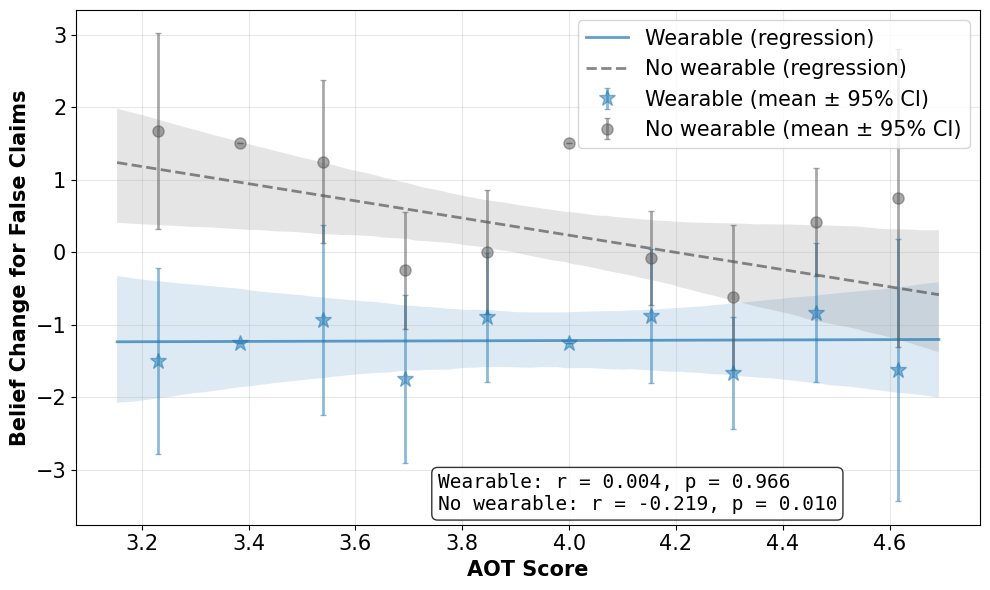}
        \vspace{-0.5cm}
        \caption{Relationship between Actively Open Thinking (AOT) ratings and belief change (post study -- baseline; -ve belief change means reduction in belief after intervention) in the false claims. Error bars represent 95\% confidence interval around linear regression.}
        \label{fig:aot}
        \Description{The scatter plot compares belief changes for false claims against AOT (Actively Open-Minded Thinking) scores, divided into two categories: Wearable and no wearable. The vertical axis is labeled Belief Change for False Claims, ranging from -3 to 3. The horizontal axis is labeled AOT Score, spanning 3.2 to 4.6. A solid blue line indicates the regression for wearable, which shows a negative trend. A dashed gray line represents the regression for No wearable, which has a seemingly neutral slope.}
\end{figure}}{
\begin{figure*}[h]
    \centering
    \begin{minipage}[b]{0.48\textwidth}
        \centering
        \includegraphics[width=\textwidth]{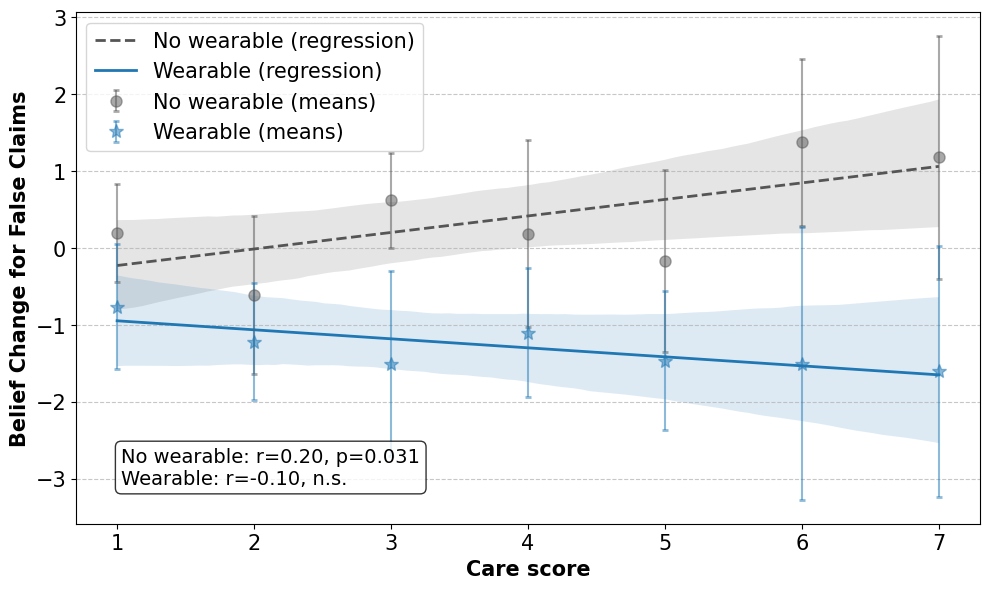}
        \vspace{-0.5cm}
        \caption{The effect of topic care on belief change (post study -- baseline; -ve belief change means reduction in belief after intervention) for false claims, by condition. Error bands represent 95\% confidence intervals around linear regression.}
        \label{fig:topiccare}
        \Description{The line graph displays the relationship between care scores and belief change for false claims, comparing groups with and without wearables. The x-axis represents the care score, ranging from 1 to 7, while the y-axis represents the belief change for false claims, from -3 to 3. There are two sets of data: one for no wearable and another for wearable. The no wearable shows a slight positive slope, while the wearable shows a slight negative slope.}
    \end{minipage}
    \hfill
    \begin{minipage}[b]{0.48\textwidth}
        \centering
        \includegraphics[width=\textwidth]{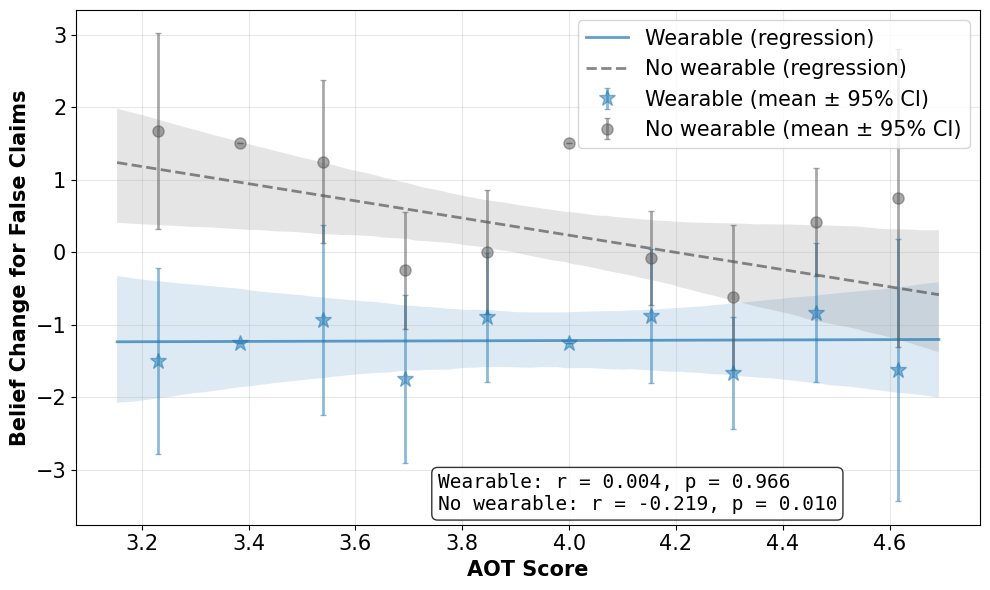}
        \vspace{-0.5cm}
        \caption{Relationship between Actively Open Thinking (AOT) ratings and belief change (post study -- baseline; -ve belief change means reduction in belief after intervention) in the false claims. Error bars represent 95\% confidence interval around linear regression.}
        \label{fig:aot}
        \Description{The scatter plot compares belief changes for false claims against AOT (Actively Open-Minded Thinking) scores, divided into two categories: Wearable and no wearable. The vertical axis is labeled Belief Change for False Claims, ranging from -3 to 3. The horizontal axis is labeled AOT Score, spanning 3.2 to 4.6. A solid blue line indicates the regression for wearable, which shows a negative trend. A dashed gray line represents the regression for No wearable, which has a seemingly neutral slope.}
    \end{minipage}
\end{figure*}}

\subsection{Wearable Intervention Corrects Beliefs Consistently Across a Range of AOT levels}
We did not find statistically significant evidence that individuals with higher AOT scores responded more strongly to the wearable intervention. As shown in Figure \ref{fig:aot}, in the wearable condition, there was no correlation between AOT and belief change ($r = 0.004, p = 0.966$). However, in the no wearable condition, a significant negative correlation emerged ($r = -0.219, p = 0.010$), indicating that without the wearable, individuals with higher AOT tended to think critically and reduce their belief in false claims, as the belief change tended to negative values for higher AOT individuals. On the other hand, individuals with lower AOT ratings showed an increase in belief in false claims. Thus, the wearable nudge effectively compensated for individual differences, creating a reduction of belief in false claims across all AOT levels.

One important methodological note is that the AOT ratings of all participants in our study fell above 3 on a scale of 1 to 6. Consequently, our findings primarily reflect the intervention's effect on individuals with moderate to high levels of analytical thinking. The observed null relationship between AOT and belief change in the wearable condition may not generalize to populations with lower AOT scores, who might respond to the nudge differently. This result highlights the need for future research to re-evaluate this relationship. 

\subsection{Prior Exposure and Belief Change}
\begin{figure}
    \centering
    \begin{minipage}[b]{0.49\textwidth}
            \includegraphics[width=\linewidth]{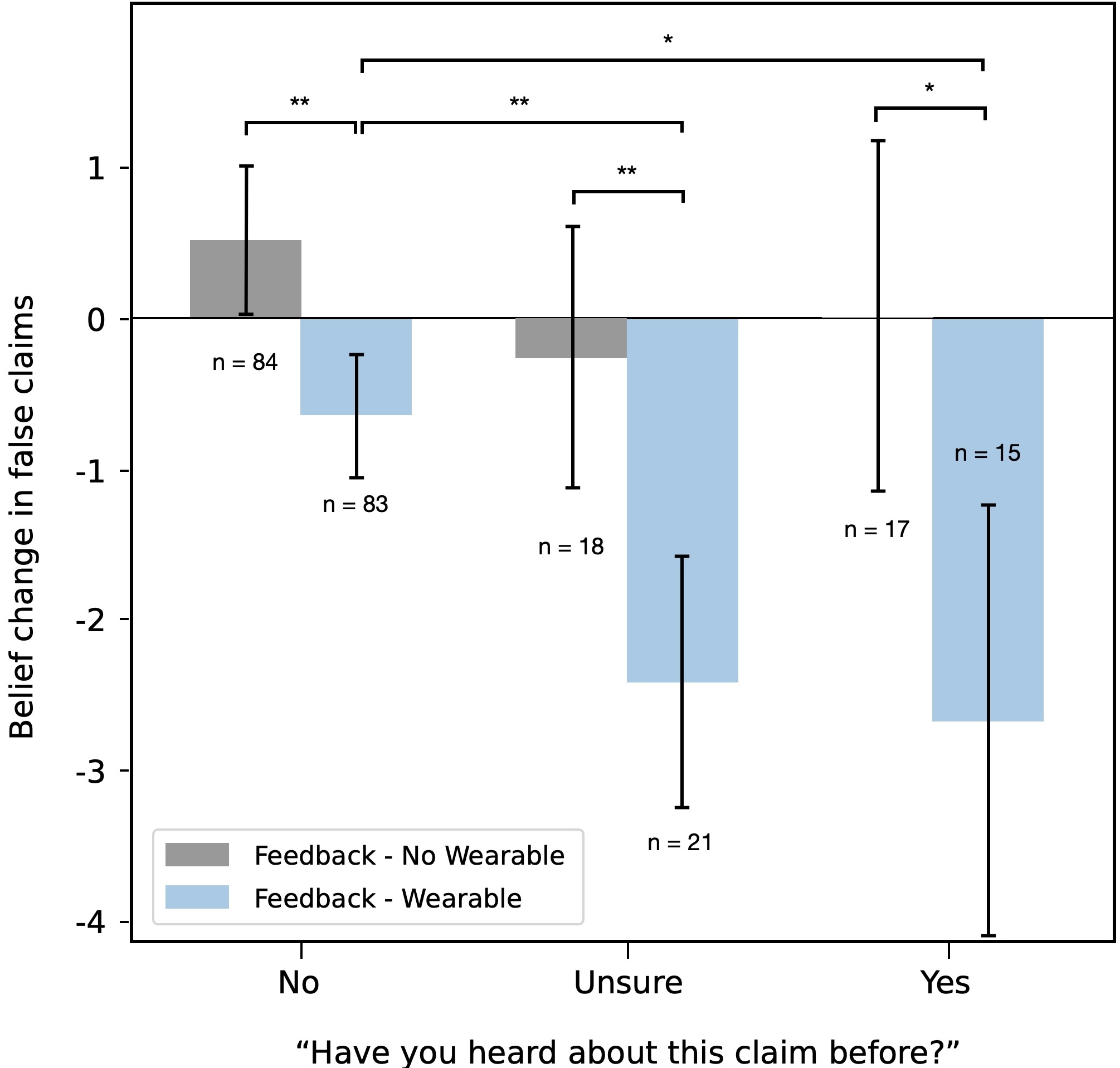}
    \caption{Relationship with prior exposure to the statements and belief change (post intervention -- baseline; -ve belief change means reduction in belief after intervention) in false claims. Error bars represent 95\% confidence interval. ``n'' is the number of instances in each prior exposure category under each condition. *p<0.05, **p<0.001, ***p<0.0001 --  Parametric t-test or non-parametric Mann-Whitney U test were conducted to assess statistical significance of differences on the pairs depending on whether the distribution of the difference was normal or not.}
    \label{fig:priorexposure}
    \Description{The bar graph illustrates the change in belief about false claims based on prior exposure to those claims. The x-axis represents prior exposure to the claims with three categories: "No," "Unsure," and "Yes." The y-axis measures the belief change in false claims, with values ranging from -4 to 1. Each category has two bars: A no wearable bar and a wearable bar. For the "No" category, the no wearable bar is slightly above zero, while the wearable bar is slightly below zero. The "Unsure" category shows the no wearable bar near zero, and the wearable bar slightly lower than -2, indicating a substantial belief decrease. In the "Yes" category, the no wearable bar is slightly above zero, and the wearable bar is almost at -3, again showing a marked belief decrease.}
    \end{minipage}
\end{figure}
Figure \ref{fig:priorexposure} indicates that the wearable intervention was consistently effective at reducing belief in false claims across all levels of prior exposure (“No”, “Unsure”, “Yes”), whereas the no-wearable condition showed minimal belief change regardless of prior familiarity. Statistical tests confirmed significant differences between conditions for all exposure groups. For participants with no prior exposure (``No'') to the claims, a Mann-Whitney U test showed a significant effect 
\newtext{($U = 4638.0$, $p = 0.0002$)}. For those who were ``unsure'', an independent t-test was significant 
\newtext{($t = 3.75$, $p = 0.0006$)}. Finally, for those with prior exposure (``Yes''), a Mann-Whitney U test also revealed a significant difference 
\newtext{($U = 203.0$, $p = 0.004$)}.
Thus, the intervention successfully reduced belief in misinformation regardless of claim familiarity. Interestingly, participants with prior exposure showed a greater reduction in false beliefs compared to those without prior exposure. However, there was no statistically significant difference between the “Unsure” and “Yes” groups within the wearable condition, suggesting a similar intervention effect across these categories.

\subsection{No Additional Cognitive Load}
We found no significant difference in the subjective cognitive load experienced by participants between the two conditions. As measured by the NASA-TLX scale, the overall workload score for the non-wearable condition ($M = 2.62, SD = 1.20$) 
\newtext{did not differ significantly from} that of the wearable condition ($M = 2.40, SD = 0.87$). Given the non-normal distribution of the data (Shapiro-Wilk $p = 0.036$), a Wilcoxon signed-rank test was conducted, confirming no significant difference between conditions ($W = 171.5, p = 0.209$). This indicates that the addition of the real-time haptic fact-checking nudge did not impose additional cognitive burden on users compared to the traditional fact-checking experience. The wearable intervention thus successfully provided corrective information without overwhelming users' cognitive capacities, suggesting its potential for seamless integration into daily information consumption.

\subsection{Overall Modeling Analysis}
\newtext{
To examine how analytic thinking (AOT), care, prior exposure, ground truth, and wearable-delivered nudges jointly influenced belief updating, we fit a single linear mixed-effects model across all trials, with participant included as a random intercept. The dependent variable was \textit{change in belief} (post-intervention minus baseline), such that positive values indicate increased belief in the claim after the intervention. As nudges could only occur in the wearable condition, we modeled their effect using an interaction term between experimental condition and nudge (condition $\times$ nudge). This specification separates the baseline effect of wearing the device (condition) from the incremental effect of receiving a nudge while wearing it (condition $\times$ nudge). We additionally included the ground truth of each claim (true vs.\ false) to account for systematic differences in how participants updated beliefs about true versus false statements.
}


The model (Table~\ref{tab:mixedlm_main}) revealed that belief change was significantly influenced by care, prior exposure, claim ground truth, and the interaction between condition and nudge. Higher care was associated with greater belief in information to be true ($\beta = 0.10, p = .023$), regardless of them actually being true or false. Prior exposure substantially reduced belief change ($\beta = -0.69, p < .001$), indicating that participants were less likely to revise their beliefs about claims they had encountered previously. Ground truth was a strong predictor, with significantly more belief reduction for false claims than for true claims ($\beta = -1.03, p < .001$). Experimental condition alone (wearable vs.~non-wearable) had no effect on change in belief ($\beta = 0.02, p = .91$), indicating that simply wearing the device did not influence judgments. In contrast, nudges delivered via the wearable had a large and highly significant effect, as reflected by the condition $\times$ nudge interaction ($\beta = -1.87, p < .001$). Because nudges primarily occurred for false claims, this interaction indicates that wearable cues strongly reduced belief in misinformation. AOT showed a marginal negative association with belief change ($\beta = -0.43, p = .062$). 

However, this overall model is limited in its ability to analyze predictors by claim type. For example, while higher care was associated with an increase in belief, it cannot determine whether this increase pertains to true or false claims. We therefore provide a more detailed, claim-wise analysis in Appendix A.

\begin{table}[t]
\centering
\caption{Linear mixed-effects model predicting belief change across wearable and non-wearable conditions. Participant was included as a random intercept. Positive coefficients indicate increased belief in the claim ($\beta$ = coefficient, SE = Standard Error, $z$-statistic = $\beta$/SE).}
\label{tab:mixedlm_main}
\begin{tabular}{lrrrr}
\toprule
\textbf{Predictor} & \textbf{$\beta$} & \textbf{SE} & \textbf{$z$} & \textbf{$p$} \\
\midrule
Intercept            & 2.99 & 0.92 & 2.26 & .001 \\\hline
Care                 & 0.101 & 0.045 & 2.27 & .023 \\\hline
AOT                  & -0.425 & 0.227 & -1.87 & .062 \\\hline
Prior Exposure       & -0.685 & 0.112 & -6.10 & $<.001$ \\\hline
Condition (Wearable=1,\\ Non-wearable=0) & 0.024 & 0.209 & 0.12 & .908 \\\hline
Condition $\times$ Nudge & -1.870 & 0.283 & -6.61 & $<.001$ \\\hline
Ground Truth\\(True=0, False=1)  & -1.028 & 0.201 & -5.11 & $<.001$ \\
\midrule
Group Variance & 0.032 & .044 \\
\bottomrule\\
\multicolumn{5}{p{\linewidth}}{\footnotesize
\textit{Model summary:} $N = 544$ observations; 34 participants; min/max observations per participant = 16/16; REML estimation; log-likelihood $=-1137.13$; residual variance (scale) = 3.74; model converged.
}
\end{tabular}
\end{table}

\subsection{Qualitative Analysis}
We conducted reflexive thematic analysis \cite{braun2021one} on open-ended responses (Section~\ref{sec:qual}), iteratively refining codes to derive five dominating themes. Analysis involved familiarization, initial coding,  theme development, review, definition, and reporting. \newtext{Three authors independently engaged with the data and then met for collaborative discussions to develop and refine themes, and document analytic decisions.} 
The resulting five themes are discussed below:
\begin{enumerate}[leftmargin=15pt]
    \item \textbf{Verification Support \& Reduced Effort:} \\
    Participants emphasized that the wearable fact-checker provided tangible support in verification and reduced the effort otherwise required. Without the wearable, participants described fact-checking as a cognitively demanding, time consuming and often \textit{``frustrating''} (P6) process. For example, 
    \begin{quote}
        \textit{``It takes time to check the information from google and chatgpt. So I will be lazy to do fact check and choose to believe what the videos teach us.''} (P11)
\end{quote} 
Others admitted that they often avoided verification altogether unless the topic was highly important\newtext{, such as health-related claims (P30)}. 
In contrast, the wearable was perceived as a low-friction aid that made the experience of fact-checking faster and easier. For example,
\begin{quote}
    \textit{``Such a system could actually act as a truth companion, which will, definitely take off my precious time spending, searching for stuff.''} (P20).
\end{quote} 

\item\textbf{Enables Reflection, Triggers Verification, \& Shifts Perceptions: }\\Several participants described that the nudges acted as prompts for \textbf{curiosity} and \textbf{reflection} \newtext{  making them question and verify claims more often (P8, P11)}. 
Others felt that the system served as a gentle \textbf{reminder} to verify \newtext{casual information} they might otherwise overlook \newtext{(P22)}.

Participants also reported becoming more \textbf{engaged} and \textbf{attentive} while watching. 
\begin{quote}
    \textit{``I waited for the `lie detector' to buzz before to pass the video''.} (P4)
\end{quote}
\begin{quote}
\textit{``The tool did help me to stay alert, especially when the notification came.''} (P32)
\end{quote}

This contrasted with descriptions of natural media consumption habits without the wearable, where participants admitted to passive listening \newtext{(P32),} 
and being \textit{``lazy to check''} (P30) or \textit{``forget to check''}(P28).

The wearable's \textbf{subtlety} was highlighted as a strength as it provided awareness without being obtrusive \newtext{(P30)}. 

Participants' reactions varied depending on whether the wearable confirmed or challenged their initial judgments. Some described \textbf{reinforcement} of their initial judgment \newtext{(P22)}, 
others reported \textbf{genuine shifts in their perception} toward critical thinking \newtext{after the watch buzzed, such as from a neutral opinion to a more critical opinion (P19), or stopping absorbing the content (P30).} 
A few even described \textbf{productive tension} when their instincts clashed with the wearable: 
\begin{quote}
    \textit{``Like, for example, if I'm very strongly believing that wearing socks is going to improve my sleep, right? And the thing is buzzing. That would annoy me enough to prove the watch is wrong, and then I'll go and check it somewhere else. So that contradiction actually makes you reflect more''}. (P30)
\end{quote}
These qualitative findings support our hypothesis H2 that the nudges promote verification activities.
\item \textbf{Trust \& Reliance}:\\
Participants expressed ambivalence toward trusting the wearable. Many were \textbf{cautious}, noting that AI-driven outputs could be inaccurate \newtext{(P32)}. 
 Several emphasized that \newtext{the information provided on the watch was too little, and} without \textbf{transparent} sourcing, the watch's prompts lacked credibility \newtext{(P12)}. 
 Others worried about omissions, questioning whether \textbf{missed prompt}s might cause them to accept misinformation by default (P29).

At the same time, the wearable sometimes triggered \textbf{over-reliance}. A few participants admitted they began to defer entirely to the device \newtext{(P6)}, 
or waited passively for a buzz before judging a claim (P11). This prompted self-reflection later, with some feeling that they were \textit{“too trusting”} (P11) towards the buzzes.

A few participants also noticed the errors (intentionally simulated AI errors) the system was making, which made them suspicious:
\begin{quote}
    \textit{``I noticed is that it didn't buzz when a guy was trying to tell me he invented a car that ran on RF waves, which as someone with a physics background, that is absolute nonsense. Yeah, it made me think, oh, how come it hasn't [buzzed for] this?''} (P29) 
\end{quote}
So despite over-reliance on the false nudges, the nudge omission errors didn't change the participants' belief (Figure \ref{fig:h2_results}), but \textbf{reduced their trust} in the system.

Trust was also described as evolving over time. Initial skepticism often gave way to \textbf{conditional acceptance} as participants cross-checked prompts and found them accurate \newtext{(P11)}. 
Others anticipated that \textbf{repeated consistency} would strengthen reliance, though rarely to full confidence \newtext{(P32, P34)}. 
Participants suggested that trust could be strengthened through transparency and references, such as linking to \textbf{credible sources} (P22, P29). Yet even with such features, they highlighted that ultimate verification would remain a personal responsibility.

\item \textbf{Situating Use in Everyday Life:}\\
Participants emphasized that the value of the wearable was highly situational. Many envisioned using it in formal or high-stakes contexts, such as business meetings, political debates, or sales pitches, where rapid verification could support credibility and decision-making \newtext{(P1, P2, P13, P18, P20, P33, P34). For example, }  
\begin{quote}
    \textit{``If I walk to a booth and someone makes bold claims about their product, then that’s very useful for me … I don’t trust salesmen.''} (P34)
\end{quote}
 Others highlighted everyday scenarios like scrolling through social media (P3) or for seniors less inclined to verify information themselves (P30). At the same time, participants identified clear boundaries of use. Some considered it unnecessary or inappropriate in trusted contexts, such as medical consultations (\textit{``I won’t use it when I go to the hospital. I trust the doctor’s profession''} (P33)), or impolite in personal conversations (P27). For interest-driven topics, participants preferred to rely on their own judgment\newtext{, rather than relying on the watch (P2)}. 
 Others noted it could feel burdensome in tasks demanding deep focus, such as literature reviews (P33). Ultimately, participants framed the wearable not as a universal tool but one whose utility depended on the stakes of the interaction, the credibility of the source, and the social norms of the setting.

\item \textbf{Design Considerations}:
\end{enumerate}
Participants offered a range of suggestions to improve usability and acceptability of the wearable fact-checker, highlighting issues of interface clarity, personalization, and control. Many found the current notifications too brief, and wanted richer content and links to sources so they could verify further\newtext{, for example, a historical log to revisit later (P6)}. 

Some emphasized alternative modalities, such as custom vibration patterns or even audio cues. Some also wondered if a smartglass would be a better alternative to a smartwatch\newtext{, as it won't need one to look away during conversations and might be socially acceptable (P28).} 


Participants also stressed the importance of user agency and control, repeatedly calling for an on/off toggle or selective use in particular contexts: 
\begin{quote}
    \textit{``I want to be able to turn it on/off because it can be distracting.''} (P1)
\end{quote}
\begin{quote}
    \textit{``I would turn it on when I was watching the news, or during project discussions.''} (P32)
\end{quote}
 Agency was tied to the idea that rather than being constantly \textit{“always listening,”} they preferred a system that respected user choice (P30, P5).

Finally, participants raised social and privacy concerns. Several felt the device would be awkward in interpersonal interactions \newtext{for example, feeling of distrust between friends (P11).} 
Others worried about being recorded or judged in real-time conversations \newtext{(P29). }
Privacy was described as particularly sensitive given that fact-checking can involve highly personal contexts (P25, P31). Thus providing control to the user for when fact-check should be done becomes one of the key ethical guardrails.


\section{Discussion}
\subsection{Wearable Nudges as Assistive Accuracy Prompts (H1)}
Our findings confirm H1: subtle haptic alerts on the wrist nudged users into more deliberate processing, reducing belief in false claims without undermining true ones. This pattern aligns with dual‐process accounts of reasoning: the wearable's alert appears to shift users momentarily from intuitive (System 1) judgment to analytic (System 2) scrutiny \cite{kahneman2011}. In effect, the haptic cue acted like a just-in-time accuracy prompt \cite{pennycook2022accuracy}, triggering a brief accuracy mindset that improved discernment. Prior work has shown that asking people to think about accuracy increases the quality of news sharing by reducing belief in false headlines \cite{pennycook2022accuracy}. Our device achieved a similar outcome, i.e.~repeated misinformation is typically more believable due to processing fluency \cite{hassan2021effects}, but the wearable’s timely buzz interrupted this “illusory truth” effect and reminded users to double-check. Importantly, belief in true information was preserved, suggesting that the device did not induce undue doubt in true content. In other words, the system serves as assistive augmentation \cite{tan2025assistive} by extending users’ cognitive capacity for truth discernment, i.e.~it amplifies reflective thinking at the right moment without replacing human judgment.

\subsection{Just-in-Time Support Encourages Verification (H2)}
H2 was also supported: participants in the wearable condition performed significantly more fact-checking activities than in the no wearable condition. The haptic alerts prompted users to pause and seek evidence, indicating increased engagement. This effect fits the paradigm of just-in-time adaptive interventions (JITAIs): the device provided support “at the moment and in the context that the person needs it” \cite{intille2004ubiquitous,nahum2016just}. As the system simulated continuous monitoring of the audio content and only buzzed for potentially dubious claims, it delivered help precisely after the claim occurred, targeting the timing gap between the user hearing a claim and forming a judgment. According to JITAI theory, this kind of contextual timing maximizes impact, which likely explains the higher verification behavior (H2) without burdening users otherwise. In line with this, our measures showed no increase in cognitive load in the treatment condition compared to control condition. The minimal, glanceable feedback design (a quick vibration and optional brief on-screen cue) successfully delivered just-in-time alerts that raised curiosity and induced verification activities without overwhelming the user.

\subsection{Managing AI Errors and Trust}
AI errors (false positives and false negatives), simulated to represent a real-world AI fact-checking system inaccuracies, played a notable role in shaping belief change. False positives were especially consequential: when the watch flagged true statements as false, participants’ belief in those statements decreased. In contrast, false negatives had little effect, as participants continued to believe those unflagged false claims same as before the intervention. Given that participants were told the device relied on an AI system, their reactions reflected their underlying level of trust in AI (mean trust rating $3.97\pm1.22$ on a 7-point scale). Despite this mid-range baseline, many ended up trusting and relying on the watch’s nudges, even erroneous ones. Several interviewees described confusion when the watch contradicted their intuition, yet after initial skepticism they tended to accept its judgments more readily over time.

This pattern highlights risks of over-reliance and the need for trust calibration. Perfect fact-checking is difficult, so systems must communicate uncertainty transparently. Participants themselves suggested features such as confidence scores or source attributions to contextualize alerts. Similar to findings by Danry et al.~\cite{danry2020wearable}, explainability—through references, brief rationales, or optional logs—was seen as essential to building trust --- but can also lead to overreliance if not calibrated correctly \cite{danry2023don}. Recently, Kim et al.~\cite{kim2025fostering} showed that when sources are provided, the reliance on erroneous responses reduces. Thus, designing wearable fact-checkers with transparent, contextual feedback could temper over-reliance while preserving the benefits of timely accuracy nudges.

\subsection{Design Implications and Ethical Considerations}
Our findings point to several design implications for wearable fact-checking systems. First, the results reinforce the assistive augmentation principles \cite{tan2025assistive}: systems should extend human cognitive capacities rather than replace them. The haptic nudges in our study exemplify augmentation by scaffolding reflective thinking at the right moment, without dictating user decisions. Extending this approach, designers should consider mechanisms for calibrating trust, such as presenting confidence scores, citing sources, or offering optional post-hoc explanations. Such transparency helps users maintain agency and autonomy while benefiting from AI support.

Ethical considerations are equally important. Our results highlight the risk of over-reliance on imperfect AI judgments, which may lead users to discount true information or uncritically accept false information. This raises questions about responsibility and accountability when an AI-supported intervention influences users’ beliefs. Designers must therefore implement safeguards that (i) make system limitations clear, (ii) encourage verification rather than passive acceptance, and (iii) respect user consent and privacy, especially when continuous monitoring of conversations is involved. Positioning these systems within the broader discourse on assistive augmentation highlights the need to balance empowerment with user autonomy, trust calibration, and transparency.

\subsection{Limitations and Future Work}
This study has several limitations. 
\newtext{We acknowledge that our study does not directly isolate the benefit of the wearable form factor; rather, it evaluates the impact of real-time feedback on belief and verification behaviors, by comparing instant feedback and no feedback conditions. } \newtext{Accordingly, our contribution is primarily a behavioral and interactional one: it characterizes how just-in-time fact nudges influence belief revision and verification under idealized conditions, rather than demonstrating a deployable end-to-end system.} \newtext{Moreover, we conduct the study in a controlled and simulated environment which helped us study the impact of such an instant fact-check feedback system on human behavior, in isolation from other system-related factors such as latency. }

Importantly, 
the system was implemented in a Wizard-of-Oz paradigm: haptic buzzes were pre-scripted to align with curated video content. 
While this design isolated human responses to just-in-time nudges, it does not reflect the technical challenges of deploying a real-time system, such as \newtext{recognizing speech in noisy environments,} detecting claims in unconstrained media, verifying them accurately, and delivering feedback with minimal latency. In particular, the study implicitly assumes near-perfect speech recognition, rapid claim detection, and highly reliable fact verification capabilities similar or better than those of current state-of-the-art fact-checking pipelines. Moreover, the controlled lab context likely amplified verification behaviors: participants reported fact-checking more often than they would in everyday settings. \newtext{Watching pre-verified videos under controlled conditions also diverges from the intended real-world use case of everyday conversation.}
\newtext{As a result, our findings reflect user responses to an idealized version of fact-nudging, and therefore, it remains unclear how much of the observed effects would persist under more realistic error rates, delayed notifications, or missed detections, and whether degraded model performance would attenuate or change user behavior.}
Thus, ecological validity remains an open question, as the frequency and depth of verification may diminish when users are multitasking, distracted, or less motivated. 

Future work should therefore extend these findings to functioning end-to-end systems 
\newtext{and extend testing into semi-naturalistic dialogues or live conversational scenarios}. 
\newtext{Such systems would enable systematic ablations over component accuracy and latency, allowing researchers to quantify trust, reliance, and intervention effectiveness in a real-world setting.} 
\newtext{Demonstrating technical feasibility through prototype deployments will also be necessary to assess whether current models can support meaningful real-world impact.}

\newtext{While we focus on smartwatches, these insights are not inherently tied to wearables and may generalize to smartphones, smartglasses, or other head-worn devices and desktop interfaces that support timely, glanceable interventions.}

Another key direction is examining the longitudinal impact of wearable nudges. Our study focused on short-term interactions, but repeated exposure may shape trust calibration, reliance patterns, and sustained verification habits over time. Just-in-time adaptive intervention (JITAI) frameworks \cite{intille2004ubiquitous,nahum2016just} emphasize that effective support depends not only on momentary timing but also on adaptation to long-term behavioral trajectories. Understanding whether wearable nudges maintain their efficacy, fade into the background, or risk habituation across extended use is necessary for establishing durable benefits. The participant pool was also limited in size and demographics, constraining generalizability. Future research should explore more diverse populations, adaptive trust calibration strategies, and richer modalities of feedback beyond haptic and text cues.


\section{Conclusion}
Our study validates that a wrist-worn fact-checker can act as an assistive augmentation \cite{tan2025assistive}, improving truth discernment in line with dual-process theory \cite{kahneman2011} and accuracy-prompt research \cite{pennycook2022accuracy}. The just-in-time, peripheral design resulted in more verification activities while avoiding extra cognitive burden. At the same time, user feedback highlights the importance of trust, explanation, and control. These insights inform a design path toward transparent, user-centered wearable truth aids. Future work should implement these recommendations and evaluate how sustained use shapes real-world information consumption and belief accuracy.

\begin{acks}
This research is supported by NUS Innovation \& Venture Creation Award, as well as by the National University of Singapore Start-Up Grant Scheme.
\end{acks}

\bibliographystyle{ACM-Reference-Format}
\bibliography{sample-base}

\appendix
\section*{Appendix}
\input{appendix_cameraready}

\end{document}

%% file: appendix_cameraready.tex

\begin{table*}[h]
\centering
\caption{List of video clips used in the study, with fact-check verdicts and explanations.}
\begin{tabular}{p{0.35\linewidth}p{0.09\linewidth}p{0.2\linewidth}p{0.3\linewidth}}
\toprule
\textbf{Claim (excerpt)} & \textbf{Verdict} & \textbf{Topic} & \textbf{Explanation} \\
\midrule
\multicolumn{4}{l}{\textbf{Video 1}} \\
Robert Downey Jr. has reportedly signed his contract to return to the MCU. & \textbf{\textcolor{green!60!black}{TRUE}} & Public Figures & Robert Downey Jr. did not sign a contract to return to the MCU. \\
A bear ripped a hole in a woman's screen to use her swimming pool in Florida. & \textbf{\textcolor{green!60!black}{TRUE}} & Animals & Not applicable. \\
Wearing socks to bed increases blood circulation to your feet. & \textbf{\textcolor{green!60!black}{TRUE}} & Health & Not applicable. \\
Manatees can affect their buoyancy through farting. & \textbf{\textcolor{green!60!black}{TRUE}} & Animals & Not applicable. \\
22-year-old Alyssa Carson will be sent to Mars in 2013 and never return. & \textbf{\textcolor{red}{FALSE}} & Public Figures & She is not part of NASA's astronaut team. \\
Peanut butter and jelly actually adds 33 minutes and 6 seconds to your life. & \textbf{\textcolor{red}{FALSE}} & Health & Extends Health-Adjusted Life Expectancy (HALE), not actual lifespan. \\
British tanks are equipped with a vessel to boil tea or coffee. & \textbf{\textcolor{green!60!black}{TRUE}} & History & Not applicable. \\
Max Wolchikumbuzo developed the first EV that drives without recharging. & \textbf{\textcolor{red}{FALSE}} & Technology & He did not build an electric car that never needs to be charged. \\
\midrule
\multicolumn{4}{l}{\textbf{Video 2}} \\
Every 17 years, a massive swarm of cicadas emerge from the ground; males reach 100 decibels. & \textbf{\textcolor{green!60!black}{TRUE}} & Animals & Not applicable. \\
Danish parents let their babies nap alone outside. & \textbf{\textcolor{green!60!black}{TRUE}} & Health & Not applicable. \\
Newborns can only see in black and white in their first week. & \textbf{\textcolor{red}{FALSE}} & Health & Newborns can see some muted colors. \\
Cristiano Ronaldo offered his hotel to shelter victims of the 2023 Morocco earthquake. & \textbf{\textcolor{red}{FALSE}} & Public Figures & He did not offer the earthquake victims to stay for free. \\
Cats are banned in the town of Longyearbyen, Norway. & \textbf{\textcolor{green!60!black}{TRUE}} & Animals & Not applicable. \\
"Pokémon GO" developer Niantic used player data to build an AI navigation system. & \textbf{\textcolor{green!60!black}{TRUE}} & Technology & Not applicable. \\
Canada has more lakes than the rest of the world combined. & \textbf{\textcolor{red}{FALSE}} & Geography & Canada has more lakes than any other country, but not more than the rest of the world combined. \\
Heath Ledger improvised in the hospital scene of \emph{The Dark Knight} after an explosion failed. & \textbf{\textcolor{red}{FALSE}} & Public Figures & The delayed explosions were expected, not improvised. \\
\bottomrule
\end{tabular}
\label{tab:claims}
\end{table*}

\begin{table*}[h!]
\centering
\caption{Overview of questionnaires administered at different stages of the study.}
\begin{tabular}{p{0.12\linewidth}p{0.22\linewidth}p{0.38\linewidth}p{0.18\linewidth}}
\toprule
\textbf{Stage} & \textbf{Construct} & \textbf{Example Item(s)} & \textbf{Scale / Response Type} \\
\midrule
\multirow{6}{*}{Initial} 
 & Demographics & Age group, gender & Multiple choice \\
 & Verification frequency & ``How often do you verify information you encounter on social media / TV news / YouTube / podcasts / short-form video / online articles / casual conversations / meetings?'' & 7-point Likert (1 = Never … 7 = Always) \\
 & Verification methods & Fact-checking sites, search engines, trusted people, platform labels, AI platforms, other & Multiple choice + open-ended \\
 & Barriers to verification & Lack of time, not sure how, trust source, forget to check, don’t care enough, agree with claim & Multiple choice (rank order) + open-ended \\
 & Trust in AI & ``To what extent would you trust information provided by an AI system?'' & 7-point Likert \\
 & Topic relatability & ``To what extent would you care to verify claims in health / public figures / animals / technology / history / geography?'' & 7-point Likert \\
 & Openness to updating beliefs & Actively Open-Minded Thinking (AOT) scale~\cite{stanovich2023actively} & 6-point Likert (no neutral option) \\
\midrule
\multirow{2}{*}{Baseline} 
 & Prior exposure & ``Have you seen or heard about this statement before?'' & Yes / Unsure / No \\
 & Belief and confidence & ``To what extent do you believe this statement?''; ``How confident are you in your belief?'' & 7-point Likert (belief: false–true; confidence: low–high) \\
\midrule
\multirow{3}{*}{Post-video} 
 & Belief and confidence (re-assessed) & Same as baseline, for the claims shown in the video & 7-point Likert \\
 & Verification importance & ``How important is it for you to verify this claim?'' & 7-point Likert \\
 & Cognitive workload & NASA-TLX~\cite{hart1988development} & Standardized scales \\
\midrule
Qualitative \& Interview & Reflections and feedback & ``Describe your experience with and without the wearable system.'' (open-ended questions + semi-structured interview) & Open-ended \\
\bottomrule
\end{tabular}
\label{tab:questionnaire}
\end{table*}

\begin{figure*}[t]
    \centering
    \includegraphics[width=\linewidth]{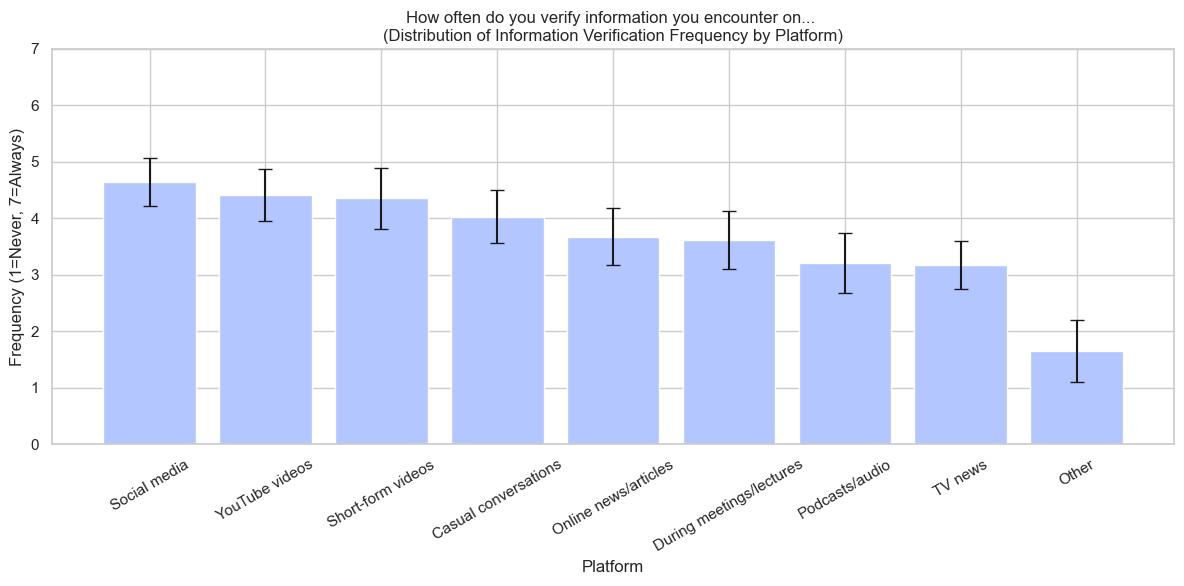}
    \caption{Response to the question: ``How often do you verify information you encounter on different platforms?''}
    \label{fig:IQ_platforms}
\end{figure*}

\begin{figure*}[t]
    \centering
    \includegraphics[width=0.8\linewidth]{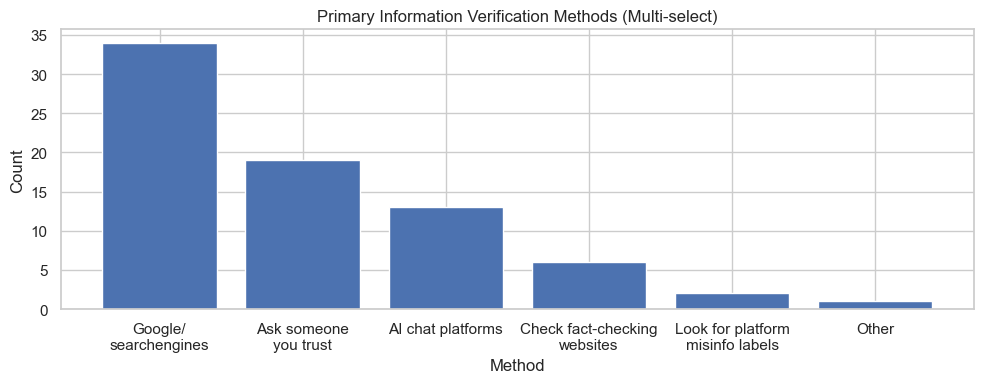}
    \caption{Response to the question ``How do you verify information?''}
    \label{fig:IQ_verification_methods}
\end{figure*}

\begin{figure*}
    \centering
    \includegraphics[width=0.8\linewidth]{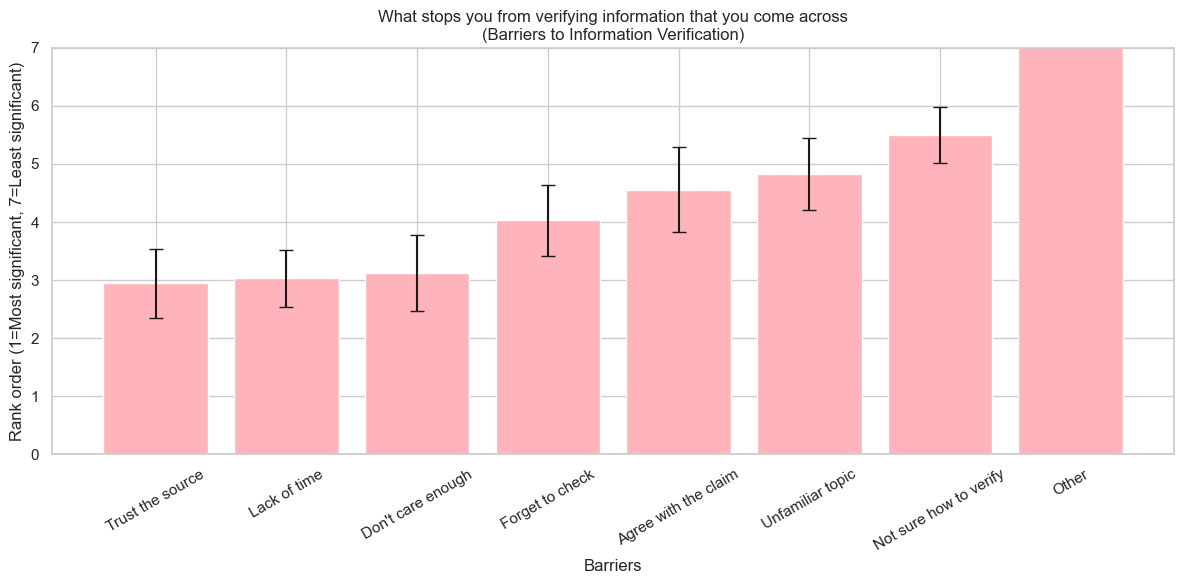}
    \caption{Response to the question ``Rank order the reasons that stop you from fact-checking (1 means the number one reason)''}
    \vspace{-8pt}
    \end{figure*}
    \begin{figure*}
    \includegraphics[width=0.7\linewidth]{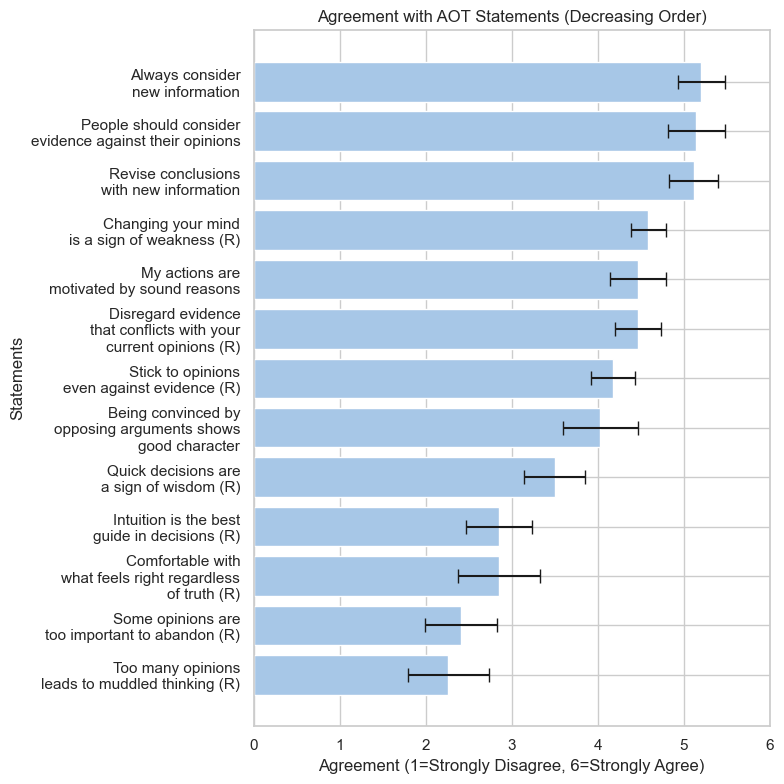}
    \caption{Response to the different questions in the Actively Open Thinking (AOT) Questionnaire.}
    \label{fig:InitialQuestionnaire2}
\end{figure*}
\section{Mixed Linear Model Regression Extended Results}
\newtext{To further examine how analytic thinking (AOT), care, prior exposure, and wearable nudges jointly influenced belief updating for each type of claim, we fit separate linear mixed-effects models for true and false claims, with participant as a random intercept. The dependent variable was \textit{change in belief} (post intervention – baseline), such that positive values indicate increased belief in the claim. Because nudges could only occur in the wearable condition, we modeled the nudge effect as an interaction term (condition × nudge), rather than including a main effect of nudge. This specification separates the baseline effect of wearing the device (condition) from the incremental effect of receiving a nudge while wearing it (condition × nudge). }

\newtext{\textbf{True claims.} For claims that were true, both care, prior exposure, and nudges in the wearable condition (condition x nudge) were found to significantly impact changes in belief (Appendix Table \ref{tab:mixedlm_side_by_side}(a)). Participants who reported higher care showed higher belief in true claims ($\beta = 0.15, p = .009$). 
For true claims, prior exposure was associated with more negative belief-change scores ($\beta = -0.63, p < .0001$), indicating that participants who had seen the claim before were less likely to increase their belief in it when encountering it again. 
For the true claims in the wearable condition on which a haptic nudge indicating the statement was false was delivered, was associated with substantial decrease in belief ($\beta = -2.34, p < .0001$), indicating that false-positive nudges in the wearable condition made participants more skeptical even when the claim was accurate. Neither AOT nor experimental condition (wearable vs.~non-wearable) independently predicted belief change (both $p > .72$).}

\newtext{\textbf{False claims.} For claims that were false, belief change was significantly associated by analytic thinking, prior exposure, and nudges in the wearable condition (Appendix Table \ref{tab:mixedlm_side_by_side}(b)). Higher AOT was associated with larger negative change scores ($\beta = -0.85, p = .013$), reflecting greater reduction in belief in false claims among more analytic participants. Participants who had previously encountered a false claim (prior exposure) showed larger negative belief-change scores when seeing it again ($\beta = -0.78, p < .001$), indicating that familiarity helped them correct their belief downward rather than accept the misinformation. False claims with a nudge in the wearable condition showed stronger reductions in belief ($\beta = -1.62, p = .006$) than false claims without a nudge (false negatives), indicating that the cues effectively promoted skepticism toward misinformation. Care and experimental condition did not explain additional variance in belief change for false claims (both $p > .55$).}

\begin{table*}[t]
\centering
\caption{Linear mixed-effects models (REML with participant random intercept) predicting \textit{change\_in\_belief} (post--baseline), fit separately for (a) true vs.\ (b) false claims. Positive coefficients indicate increased belief in the claim.}
\label{tab:mixedlm_side_by_side}

\begin{subtable}[t]{0.49\textwidth}
\centering
\setlength{\tabcolsep}{5pt}
\renewcommand{\arraystretch}{1.12}

\begin{tabularx}{\linewidth}{@{}l *{4}{>{\raggedleft\arraybackslash}X}@{}}
\toprule
Predictor & $\beta$ & SE & $z$ & $p$ \\
\midrule
Intercept & 1.570 & 1.100 & 1.43 & .154 \\
Care & 0.148 & 0.056 & 2.62 & .009 \\
AOT & -0.096 & 0.274 & -0.35 & .727 \\
Prior Exposure & -0.629 & 0.137 & -4.60 & $<.001$ \\
Condition (Wearable) & 0.007 & 0.216 & 0.03 & .975 \\
Condition $\times$ Nudge & -2.336 & 0.479 & -4.88 & $<.001$ \\
\midrule
Group Variance & 0.011 & 0.056 &  &  \\
\bottomrule
\end{tabularx}

\vspace{2pt}
\begin{minipage}{\linewidth}
\footnotesize
\textit{Model summary:} $N = 306$ observations; 34 participants; min/max observations per participant = 9/9; MixedLM (REML) estimation; log-likelihood $=-621.993$; residual variance (scale) = 3.35; model converged.
\end{minipage}

\caption{True claims}
\label{tab:mixedlm_true}
\end{subtable}
\hfill
\begin{subtable}[t]{0.49\textwidth}
\centering
\setlength{\tabcolsep}{5pt}
\renewcommand{\arraystretch}{1.12}

\begin{tabularx}{\linewidth}{@{}l *{4}{>{\raggedleft\arraybackslash}X}@{}}
\toprule
Predictor & $\beta$ & SE & $z$ & $p$ \\
\midrule
Intercept & 3.872 & 1.398 & 2.77 & .006 \\
Care & 0.041 & 0.069 & 0.60 & .551 \\
AOT & -0.854 & 0.343 & -2.49 & .013 \\
Prior Exposure & -0.776 & 0.188 & -4.14 & $<.001$ \\
Condition (Wearable) & -0.130 & 0.586 & -0.22 & .824 \\
Condition $\times$ Nudge & -1.622 & 0.591 & -2.74 & .006 \\
\midrule
Group Variance & 0.000 & 0.084 &  &  \\
\bottomrule
\end{tabularx}

\vspace{2pt}
\begin{minipage}{\linewidth}
\footnotesize
\textit{Model summary:} $N = 238$ observations; 34 participants; min/max observations per participant = 7/7; MixedLM (REML) estimation; log-likelihood $=-511.13$; residual variance (scale) = 4.27; model converged.
\end{minipage}

\caption{False claims}
\label{tab:mixedlm_false}
\end{subtable}

\end{table*}
